# How does your gyroid grow?: A mesoatomic perspective on supramolecular, soft matter network crystals

Gregory M. Grason[1] and Edwin L. Thomas[2]

[1]Department of Polymer Science and Engineering, University of Massachusetts, Amherst, MA 01003

[2]Department of Materials Science and Engineering, Texas A & M University, College Station, TX 77843

*ABSTRACT*

We propose and describe a framework to understand the structure of supramolecular network crystals formed in soft matter in terms of *mesoatomic* building blocks, collective groupings of amphiphilic molecules that play a role analogous to atomic or molecular subunits of hard matter crystals. While the concept of mesoatoms is intuitive and widely invoked in crystalline arrangements of sphere- or cylinder-like (micelle-like) domains, analogous notions of natural and physically meaningful building blocks of triply periodic network crystals, like the double-gyroid or double-diamond structures are obscured by the complex, bicontinuous domain shapes and intercatenated topologies of the double networks.  Focusing on the specific example of diblock copolymer melts, we propose generic rules for decomposing triply-periodic network crystals into a unique set of mesoatomic building blocks.  Based on physically motivated principles, the combination of symmetries and topologies of these structures point to mesoatomic elements associated with the nodal connections, leading to mesoatomic volumes that are non-convex and bound by smoothly curved faces, unlike the more familiar Voronoi polyhedral shapes associated with sphere- and cylinder-like mesoatoms.  We analyze the shapes of these mesoatoms, their internal structure and importantly their local packing with neighbor mesoatomic units. Importantly, we hypothesize that mesoatoms are kinetically favored intermediate structures whose local shapes and packing template network crystal assembly on long time scales.  We propose and study a minimal energetic model of mesoatom assembly for three different cubic double-network crystals, based only local shape packing, which predicts a detailed picture for kinetics of intercatenation and surface growth.  Based on these analyses, we discuss several possible extensions and elaborations of the mesoatomic description of supramolecular soft matter network crystals, most notably the implications of mesoatomic malleability, a feature that distinguishes soft matter from hard matter crystals.  We describe experimental observations of malleable mesoatomic units in the precursor sponge phase as well as in ordered cubic networks, and suggest possibilities for observing mesoatoms in primordial, pre-crystalline states.

# I. Introduction

Supramolecular assembly into long-range ordered "soft crystals" is occurs for nearly every class of soft molecular assembly, from surfactants in water [1] and liquid crystals, to block copolymers [2] and so-called "giant amphiphiles" [3]. Paradigmatic examples of this are the micellar lattice phases of amphiphilic molecules [4]. Local segregation between chemically immiscible regions on the same molecule, caused by both repulsive and attractive local forces, drive the aggregation of groupings of molecules. As mismatch between local shapes of distinct parts of the molecule, possibility in combination with affinity for solvent exposure, can favor curvature of the interface between those distinct molecular regions favoring, for example, to cylindrical or spherical micelle aggregates. In neat systems, or at high enough concentrations in solvated systems, the micellar units strongly interact [5,6], leading to the formation of periodically ordered equilibrium states. In these ordered states, micellar groupings of molecules are situated in crystalline arrangements (e.g. BCC, FCC lattices) [7–9], and hence, an individual micelle can be thought of as an *mesoatom*, analogous to the mesoatomic units that serve as the building blocks of solid state crystals.

Broadly speaking, this mesoatomic perspective has been highly valuable for two key reasons. For one, assessing the geometry mesoatomic volumes based on space-filling tessellations of the crystal packing has enabled rational frameworks for understanding thermodynamic selection of the crystal symmetry. For example, a common heuristic approach is to assume that sphere-like mesoatoms are deformed to conform to polyhedral, Voronoi-like partitions that bound the occupied Wyckoff sites and then to compare different measures of geometric distortion relative to ideal spherical shapes (e.g. minimal area) [10–14]. Rational arguments and theories along these lines predict, for example, that under certain conditions canonical simple crystal packings like BCC become unstable to surprising and much lower symmetry structures, like the Frank-Kasper crystals [15–19]. Beyond equilibrium states, the mesoatomic picture has obvious implication for kinetics and transformation pathways to soft crystal formation, in which aggregation of mesoatomic units themselves being the primary step in the hierarchical pathway for the ultimate structure formation, followed by the subsequent "binding" and rearrangement of mesoatoms into crystalline arrays taking place on much longer time scale. Hence, the structure and collective behavior of mesoatomic intermediates has a critical impact on the time scales that soft crystal structures form (i.e. nucleation and growth), as well as on the nature of defects present in the structures. Moreover, in most systems, soft crystals are formed by controlled quenches from either a lower concentration and/or higher temperature, and hence, the conditions at which mesoatomic units are "born" are in general quite different from the final state of the material, which is often solvent-free. The pathway dependence gives rise to rich possibilities for creating long-lived, essentially frozen, out-of-equilibrium states, with symmetries that are distinct from the more limited palette of purely equilibrium states.

While the picture is intuitive for periodic assemblies of convex and discretely defined domain shapes (e.g. spheres and cylinders), the basic notion of a mesoatomic unit is confounded by a whole class of supramolecular crystals, namely triply-periodic networks (TPNs), sometimes called bicontinuous phases. These are most often cubic phases that form at conditions that are intermediate to lamellar or cylindrical domain shapes [1], in which the domains themselves are continuous and topologically inter-catenated throughout the bulk structure. Two canonical examples are the double-gyroid (DG) and double-diamond (DD) phases, which in the simplest cases are composed of two types of subdomain: inner region (usually minority component) of interconnecting tubular domains (3-valent and 4-valent nodal connections for DG and DD, respectively) separated by a slab-like matrix domain, whose undulating shape roughly approximates a triply periodic minimal surface (TPMS), known as Gyroid and Diamond minimal surfaces for DG and DD, respectively [20,21]. The tubular phases of DG and DD form two inter-catenated networks: the respective tubular networks of DG and DD interlink in 10- and 6-member links (i.e. 10,3 and 6,4 nets).

The complex structures of TPN crystals are highly attractive and long-sought after for a range of functional material applications due to the high volumetric surface areas afforded by their intermaterial dividing surfaces (IMDSs) and polycontinuous domain topologies [22]. However, numerous basic questions remain about how and why they form. In the context of the mesoatomic paradigm summarized above, it is unclear what are the "elementary building blocks" of TPN crystals, since in the simplest notions of domain would imply that each bulk structure contains only *two* domains (i.e. the subnetworks) containing an extensive number of links between them. Hence, the kinetically accessible pathways that guide the combination of long-range positional order and topological domain connectivity of TPN crystals are not presently known, nor is it clear whether these correspond to characteristic groupings of molecules and how these relate to the ultimate structures.

In this paper, we propose and outline a mesoatomic construction for understanding the structure of TPN crystals of supramolecular soft matter crystals. Our discussion is primarily centered on the specific case of diblock copolymer melts as a paradigmatic example of a TPN forming system, although it can be understood that this perspective extends to other classes of soft-molecular building blocks, not to mention more complex block copolymer systems. Here, we propose some elementary principles that guide the definition of the mesoatomic building blocks of network crystals. We show how these mesoatomic units can be defined as the non-convex analogs to the Voronoi-like polyhedra that tesselate crystals composed of sphere-like domains. In the case of TPN crystals of AB amphiphiles (e.g. diblocks), mesoatoms are associated with high-symmetry, nodal regions of the network morphologies, and unlike polyhedral (e.g. Voronoi) cells, are bounded by positively- and negatively-curved (saddle-shaped) faces as well as (approximately) planar faces. We describe the structure of canonical TPN crystals, including DG, DD and double-primitive (DP) structures, in terms of the local packing of mesoatomic clusters and the topology of the networks these units form. Next, we consider a simple model in which contact between shape-complementary neighbors templates the assembly, growth and inter-catenation of the TPN crystals, and ultimately guides larger length scale morphological features such as surface faceting. We also discuss the experimental context for the observations and implications of the mesoatom concept in block copolymers. Finally, we outline a series of open questions posed by and possible extensions to the mesoatomic hypothesis for the formation and properties of TPN crystals.

## II. Defining the mesoatom

Here we outline the principles for defining and extracting the mesoatomic building blocks of TPN crystals. Our definition specializes to the case of linear AB diblock copolymer melts, as a prototypical example. In this system, ordered phases consist of uniformly filled, but molten (i.e. solvent free, but fluid) packing of brush-like subdomains of chemically distinct A- and B-type polymer chains, covalent joined at their junctions, which are localized to 2D intermaterial dividing surfaces (IMDSs) that separate the unlike brush subdomains [2,23]. In the discussion that follows, we outline possibilities and challenges to extending this paradigm to other molecular systems that form these structures. As a starting point, we assume correct knowledge of the crystallographic space group of the final crystal, the shapes and topologies of the A and B domains, and basic model of how the chains are packed in those domains. Notably, the first two of these three descriptors can be measured by careful experimentation (excepting for more realistic deviations from idealized crystallographic order), while the last is currently essentially invisible to experimental characterization. Hence, the expected chain trajectories rely on information from computational models as well as heuristic assumptions about likely chain trajectories (as discussed in detail in [24]).

To begin with, we clarify our objective in terms of the arguably much more intuitive case of crystals of spherical or cylindrical domains. In this case, one can break these structures into micellar (i.e. spherical or cylindrical) building blocks, which when packed together become warped into lower symmetry polyhedral volumes. Beyond being the structural elements of the ultimate crystal, such "micellar mesoatoms" are often understood as kinetic intermediates, forming first into spherical or cylindrical groups, and then over longer time scales, organizing into a long-range ordered and densely packed crystals [25,26]. In this context, we clarify our intended notion of *mesoatoms* of a more general class of supramolecular crystals, which are not necessarily composed of convex quasi-spherical or quasi-cylindrical molecular groupings. That is, more generally, we aim for notion of mesoatoms as groups of molecules that act collectively as "building blocks" of a supramolecular structure, and which can be identified from the ultimate crystalline structure (akin to the Voronoi-like polyhedral volumes for quasi-spherical domains, see e.g. BCC and FCC structures in Fig. 1). Ideally, the packing and deformation of mesoatoms should be useful to describe the behavior of the crystalline structure as a whole. Further, so defined mesoatoms should represent likely kinetic intermediates whose structure and organization form the basis to understand how the crystal structure forms and evolves towards its final "mature" state.

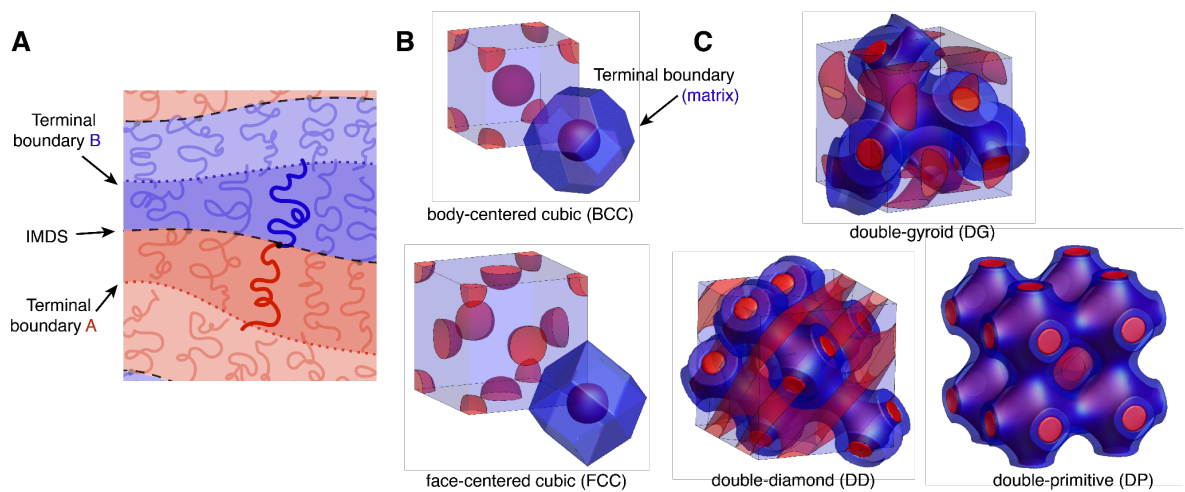

**Figure 1. Domain anatomy of block copolymer "crystal" phases –** (A) schematic illustration of subdomain packing of linear diblock copolymer chains in melt domains, where a single "domain" is defined as set of chains whose A-B junctions are associated to a particular IMDS (darker blue and red melt regions corresponding to a single domain of the quasi-lamellar geometry). (B) the putative shape of compact sphere-like domains (darkened red volumes) surrounded by blue outer volume with the expected faceting of the outer terminal boundaries separating neighboring sphere-like mesoatom domains. (C) single mesoatom domains of double-network crystals. Unlike the sphere case, due to network connectivity, there are only two domains in the entire volume (one for each network).

To understand the particular complexity of defining a useful notion of mesoatoms for TPN phases, we briefly review the anatomy of the DG crystal of diblocks, as a concrete example. The DG is cubic network morphology with $Ia\bar{3}d$ space group symmetry [27,28]. In diblock melts it is composed of two tubular network regions, usually the minority component (say the A block), separated by a matrix B block layer. The IMDSs then have the shape of a tubular surface that interconnects the 3-valent nodes. The nodes of the distinct networks, each of which can be considered a single-gyroid (SG), are centered on Wyckoff positions 16b. The *nodes* of one SG network correspond to 8 of the 16b sites (specifically these are either 8a or 8b positions of the $I4_132$ subgroup), while the 8 nodes of the second network are given by inverting the first network through the center of the unit cube. For DG, each of the SG networks is chiral, with a handedness that can be associated with the dihedral rotation between neighboring nodes [29]. For this paper, we refer to the alternate single networks of the double network crystals as "+" or "-", independent of whether the network is chiral.

As we describe below, the notion of mesoatoms is intimately connected to the concept of a *domain* in the BCP melts, following the topological definitions introduced in ref. [24]. Simply put, a *domain* corresponds to the volumes occupied by chains that have their junctions located on, or *associated to*, a particular IMDS. Hence, each domain is a type of double layer of A and B brushes, separated by the IMDS containing their common junctions. This domain decomposition, by necessity, introduces a second set of boundaries at the outer and inner "edges" of the domain, referred to as the *terminal boundaries*. The terminal boundaries are the dividing points between brush domains of the same chemistry: a segment at the terminal boundary has equal probability to associate to at least two distinct IMDSs (or IMDS positions). Colloquially, the terminal boundary can be thought of the contact surface between two opposing brush like subdomains, each of which stems from the IMDS of distinct domains. Given this notion, it has been proposed that the terminal boundary is well approximated by *medial sets* of the IMDSs [24,30,31], which are loci of midpoints within a region of the A or B component between distinct IMDS regions. In particular, for the DG network, the outer terminal surface that separates the two networks closely approximates the Gyroid minimal surface, and the corresponding boundary for DD and DP closely match the Schwarz D and P minimal surfaces respectively.[1]

In this context, the DG includes exactly 2 domains, one for each of the SG networks (which are enantiomeric), and the outer terminal boundary that separates them in the middle of the B matrix is a close approximation to Schoen's G minimal surface [32]. Note that for the crystal packings of sphere- and cylinder-like domains (e.g. Fig. 1B), each domain corresponds to a single, compact mesoatom, whereas in the double-network crystals like DG, DD and DP, only 2 constituent domains span the entire volume of the crystal. Fig. 1C shows examples of these "macroscopic" network domains for DG, DD and DP, in contrast to the compact discrete shapes of the for crystalline packings of spherical domains. As topologically well-defined objects, the 2 single-network domains of these double-network morphologies are natural groupings of molecules.

However, for the purposes of the mesoatomic construction, single-network domains are obviously problematic. First, each is macroscopically large, spanning the entire volume. And second, the two networks in the final morphology are topologically intercatenated. Simply put, there is no way for two pre-formed single network domains to interlink into the final double-network without a (kinetically prohibitive) process of a multitude of breaking and relinking events. Note that this topological problem of interlinked domains is not encountered for quasi-spherical or quasi-cylindrical domains, where each domain (corresponding to each spherical or cylindrical IMDS) constitutes a single, convex mesoatomic unit. From this perspective, it may now be intuitive to see that mesoatoms need to be defined as *pre-linked* (i.e. not yet linked) sub-elements of these single network domains. Next we ask, what is the natural and generic method to decompose the single networks into their mesoatomic constituents?

To define mesoatomic elements of TPN morphologies, we propose three basic principles:

1) The ultimate crystal structure is a symmetric, space-filling packing of many copies of a single (or at most, a few) mesoatomic motif(s)
2) Mesoatom shapes correspond to (average) volumes occupied by specific groupings of molecules (i.e. mesoatom boundaries do not cut across average chain trajectories)
3) Mesoatoms correspond to thermodynamically/kinetically favored local structures
    a) Mesoatoms possess high point group symmetry (i.e. the crystal volume includes many copies of a favored subdomain packing motif)
    b) Mesoatom dimensions should be comparable to molecular size

While the first two propositions guarantee that the ultimate structure could be "rebuilt" by assembly of the mesoatomic units, the third proposition is motivated by the notion that mesoatoms identified in the final structure can be connected to groups of molecules that are likely to pre-assemble under many growth conditions. To that end, proposition 3.a implies that packing adopts multiple copies of a favored local geometry, conferring it with a low free energy, while 3.b is requirement for the reasonably fast nucleation time to form (micelle-like) aggregates. Of course, 3.a and 3.b do not themselves guarantee that such an arrangement is kinetically favored under all conditions without careful consideration of the non-equilibrium pathways of formation, but we nevertheless argue that these principles provide a reasonable "zeroth order" framework to extract likely candidates based only on geometry, topology and symmetries of the final phase morphology, and one which can be generalize to self-assembled crystals with more complex domain topologies than spheres and cylinders.

---

[1] In general, *the inner terminal surfaces* are "weblike" structures that span the skeletal graphs [30], and while details of this shape have consequences for packing within double network phases, they are less consequential for defining boundaries of their mesoatoms.

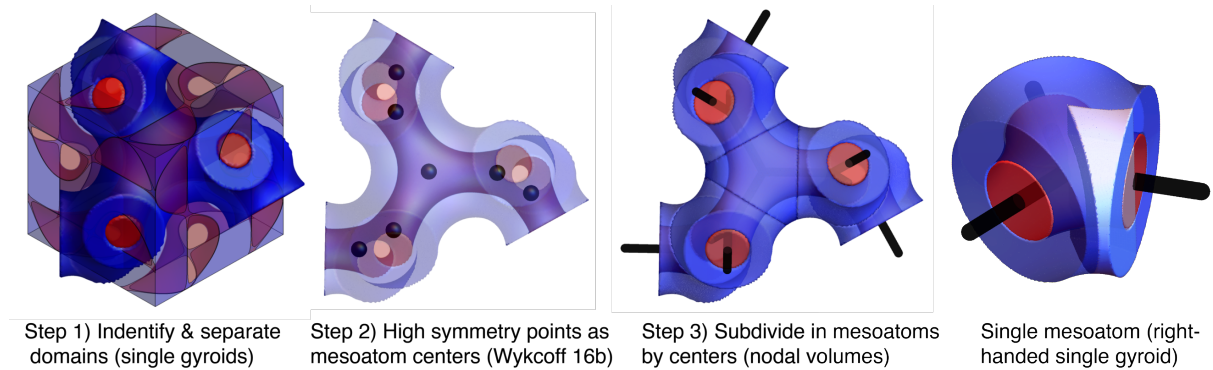

| Step 1) Indentify & separate domains (single gyroids) | Step 2) High symmetry points as mesoatom centers (Wyckoff 16b) | Step 3) Subdivide in mesoatoms by centers (nodal volumes) | Single mesoatom (right-handed single gyroid) |

**Figure 2. Mining the mesoatom –** Schematic sequence of domain to mesoatom decomposition in "+" network domain of DG. On the right, we shown the final (enlarged) mesoatom, with the black tubes highlighting the portion of the single gyroid network to which it belongs. Additionally, we highlight that this mesoatom itself is composed of 6 copies on the "asymmetric unit" (highlighted in lighter colors), that can be generated by the elements of the $D_3$ point symmetry.

Next, we describe how these physical principles can be translated into a prescription for identifying the mesoatoms from a final TPN crystal of a block copolymer melt. This approach breaks into three steps:

*Step 1) Divide the ultimate crystalline structure into individual domains (i.e. single networks)*

*Step 2) Identify the "centers" of the mesoatoms as the set of highest point symmetry positions (i.e. Wyckoff sites) fully enclosed by the single network domains*

*Step 3) Divide the single network into volumes according to the mesoatom centers*

These three steps are illustrated in Fig. 2 for mesoatoms within one of the two single gyroid domains of the DG.

In this case, the mesoatom centers correspond to Wyckoff positions 16b, which are situated at the center of 3-valent junctions (black spheres in step 2 of Fig. 2b). These positions are the intersection of three 2-fold rotation axes that meet along a 3-fold rotation axis normal to the plane spanned by the 2-fold axes, which run along the "struts" of the skeletal graph of the (10,3) network. This $D_3$ point symmetry (or .32 in Hermann-Mauguin notation) implies that there are *six copies* of the same asymmetric motif in this region. Notice that there are Wyckoff positions (24c) that sit at the centers of struts between 16b sites, but these have point symmetry ($D_2$) with fewer (4) copies, of fundamental (asymmetric) local packing motif than position 16b. (In terms of the number of copies of asymmetric unit per Wyckoff position, it is intuitive that positions with the greatest number of motif copies per site have the highest point symmetry). As we discuss in more detail below, while Wyckoff position 16a (with point symmetry $C_{3i}$) has the same number of asymmetric motifs (when accounting for point inversion) this position sits at the terminal boundary between two single gyroid domains, and as such, it can't be used to define the center of a mesoatom. We note that as consequence of these rules and the particular symmetry of the DG crystal, we find that DG mesoatoms split into two opposite chiralities, consistent with their centers at noncentrosymmetric site 16b.

In Fig. 2, we show the last step to divide the gyroid domains according to their centers (at 16b sites) into mesoatoms. Strictly speaking, proposition 2 above requires this dividing surface avoid crossing through mean molecular trajectories (see Discussion and Fig. 15 below). Specifying such a surface, of course, requires some detailed knowledge about the mean trajectories of those molecules, such as could be provided via BCP tessellations used in strong-segregation theory calculations [30,33]. However, to a good approximation, chains along the struts between nodes of the DG (or other networks) tend to radiate roughly normally the so-called skeletal bond that connects between the node centers (a more accurate picture has trajectories extending from 2D web-like terminal surfaces [24]). Hence, for our purposes, we approximate surfaces that divide between neighbor mesoatoms as planes normal to those skeletal bonds (i.e. local chain trajectories are assumed to be parallel to those dividing planes). Partitioning the network into mesoatoms of equal volumes is performed by perpendicular bisecting plane through the struts of the single gyroid separating two neighboring 16b sites.

The result of this process is shown in Fig. 2, with two sets of mesoatoms required for the DG assembly. Each mesoatom is chiral, deriving from one of the two enantiomeric single gyroid domains, and inherits the $D_3$ symmetry of the 16b sites, as well as 1/16 of the volume of the cubic $Ia\bar{3}d$ unit cell. Notably the non-convex shapes of these DG mesoatoms (as well as the counterparts for DD and DP) are more complex that the polyhedral mesoatomic shapes expected from simple crystals of spherical domains (e.g. BCC and FCC). The obvious distinction is that mesoatom volumes form DG and other TPN are bounded by two types of surfaces: negatively, curved and approximately minimal surface faces; and roughly planar faces derived from subdivision of the single network domains into high point symmetry objects. These two types of surfaces, correspond to "contact" between mesoatoms of two distinct types: inter-network (the saddle faces) and intra-network (quasi-planar strut faces). In the following sections, we analyze the shapes of these complex mesoatomic "particles," their local packing geometry in the DG, as well as their counterparts in DD and DP.

### III. Anatomy and packing of mesoatoms: DG, DD and DP

We now describe the geometric features of mesoatoms defined by the decomposition principles mentioned above for the three canonical cubic double-networks: DG, DD and DP. For the purposes of modeling the principal shape characteristics, we model the terminal surfaces dividing the two single networks domains using the single-Fourier mode level set approximation to the Gyroid, Diamond and Primitive TPMSs [34,35]. Given a diblock copolymer domain, and in particular, an IMDS for the structure, better approximations to this terminal boundary could be constructed from the medial surface in the matrix domains. Again, we make the assumption that mean chain trajectories are well-approximated as normal to the skeletal bond between two nodes (i.e. mesoatom centers), such that the faces that divide two neighbor mesoatoms within the same network are planar and perpendicular to those bonds (struts). We expect these approximations are sufficient to capture the primary structural mesoatomic motifs of distinct structures, although one should also expect at least subtle variations in the detailed shapes of both the inter- and intra-network faces shapes, not to mention questions about temporal evolutions, thermal fluctuations and distortions away from any idealized shape, a point that we return to in discussion (see Fig. 15).

Fig. 3 shows the mesoatomic units for the three cubic double networks, highlighting the topology of distinct networks with black or white struts that protrude through the strut faces according to whether the units belong to the "+" or "-" network domains. Additionally, a view of mesoatoms whose centers belong with a cubic repeat of the structures (the translational symmetry of DG and DP are body-centered, while DD is primitive). Very briefly, the shapes of these mesoatoms are constructed following the three steps defined above, using a (single Fourier mode) level surface approximation [34,35] for the minimal P, D and G surfaces (for DP, DD and DG respectively) as model of the matrix terminal boundary that divides the double-networks into two continuous domains.

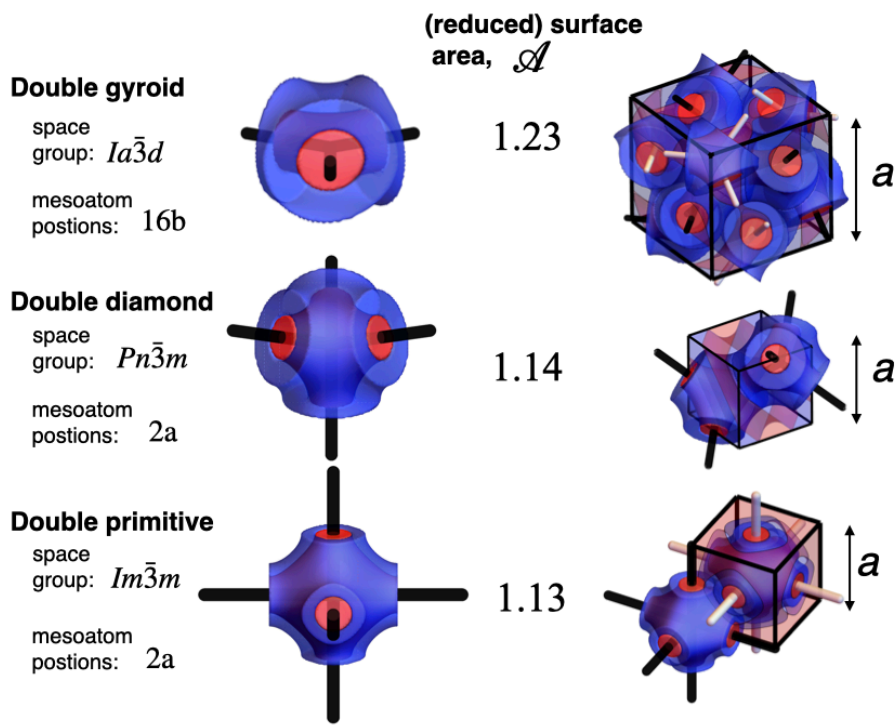

**Figure 3** – Comparison of mesoatoms of cubic DG, DD and DP. The right-most column highlights volumes of mesoatoms whose centers lie within an elementary cubic repeat of each crystal. The struts emanating from the faces of the mesoatoms are colored black and white to indicate which of the two networks they belong two (referred to as "+" or "-" networks in the text).

Notably, like the DG mesoatom introduced above, for DD and DP, mesoatoms are centered on the "nodal" centers of the single network, corresponding to sites of tetrahedral or octahedral coordination, respectively. Additionally, mesoatomic shapes are bounded by two types of surfaces: saddled-shaped surfaces that divide between neighbors on different networks; and planar faces the divide between neighbors on the same network. We note that the DG structure is composed of two sets of chiral mesoatoms, and that chirality is reflected in shape of the saddle "skin", closely following the chiral shape of an oriented Gyroid surface [32].

Given the mesoatom shapes extracted from the final morphology, it is straightforward to analyze and compare their basic geometry. As non-convex volumes, it is intuitive that such shapes are bounded by relatively large surface area. Indeed, the dimensionless surface-to-volume ratio (or equivalently the isoperimetric quotient) is commonly invoked in either heuristic or geometric theories for symmetry selection in crystals of quasi-spherical mesoatom domains. Here we measure this by

$$\mathcal{A} \equiv \frac{(\text{area})}{[36\,\pi\,(\text{volume})]^{2/3}}$$

Which is the ratio of the bounding surface area (including strut and saddle boundaries) to the area of an equal volume sphere, a dimensionless number that is strictly $\geq 1$. For comparison, for the (mostly) convex, polyhedral cells for sphere packings, like BCC or even more complex Frank Kasper variants, $\mathcal{A}$ is the in the range 1.09 - 1.1 [12,19], in this sense, roughly 10% more area than spherical volume. The dimensionless area values for mesoatoms of cubic double networks are shown in Fig. 3, indicating increases of the surface area in excess of roughly 5-12% higher than the mesoatomic elements of spherical domain crystals, an intuitive consequence of the non-convex shapes of network mesoatoms. The hexavalent DP mesoatom with the highest point group symmetry ($O_h$) and whose shape is similar to a truncated octahedral shape of the BCC Voronoi cell, is lowest amongst the network mesoatoms. Network meosatoms with the lowest point group symmetry, trihedral DG mesoatoms, exhibit a substantially larger area.

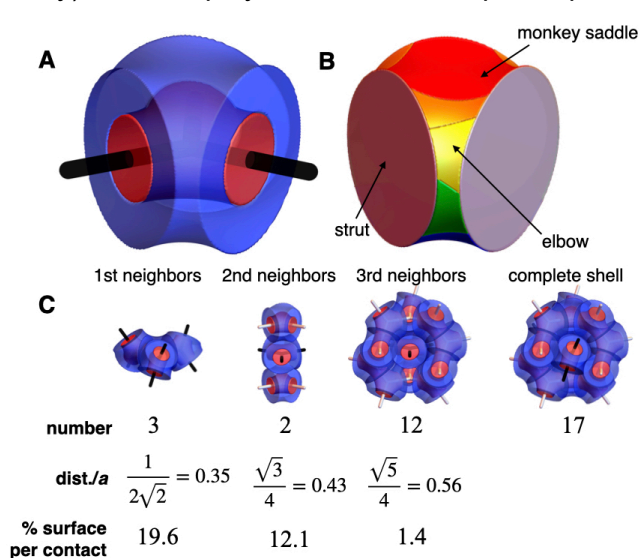

**Figure 4 –** Shape and packing of DG mesoatoms. (A) shows one of the elementary DG mesoatoms which is recolored in (B) in terms of the regions of contact between the 17 neighboring mesoatoms that touch its surface. These neighbors fall into 3 distinct sets according to their center-to-center distance and surface contact as detailed in (C).

Beyond the basic geometry of individual shapes, the mesoatomic decomposition provides valuable insights into the local packings of contacting-neighbor units, which differ considerably between the network types. We begin by describing local packing of a DG mesoatom and its surrounding, first shell of neighbors, corresponding to the set of mesoatoms sharing contact with surface of a central particle. As shown in Fig. 4, DG mesoatoms have 17 contacting neighbors, which are classified into three sets according to the center-to-center distances. A DG mesoatom has three nearest neighbors, which are strut neighbors, belonging to the same single gyroid network (and hence have the same chirality) whose centers belong to the common plane of the 3 two-fold axes. The remaining 14 neighbors belong to the other gyroid network, and are in contact with the central mesoatom along its saddle surface. In this set, there are two next nearest neighbors, which are situated "above" and "below" the 3-fold axis of the central particle (i.e. stacked along a <111> direction). These stacked pairs nestle along minimal, monkey-saddle shaped regions. The remaining 12, third nearest neighbors contact the central particle along the "elbow" regions of the saddle surface that span between two struts.

In Fig 4B, we color the surface of the DG mesoatom according to these regions of local contact, and in Fig. 4C, we give the fraction of the surface that is contacted by a neighbor of each of these types. In this way we see that the majority of the neighbor contact (57.6%) is composed of like-network strut neighbors. The remaining fraction is split between the two monkey-saddle neighbors (24.2%) and the twelve elbow neighbors (16.8%). Not unlike the better known case of Voronoi polyhedra, here we also find that contact area decreases with neighbor separation (center-to-center). Due the very unequal distributions of contact areas between the two populations, intra-network contacts dominate the surface coverage of DG mesoatoms, despite the fact that they are overwhelmed in number 14:3 by inter-network contacts. Below we consider the potential ramifications of the contact area distributions between neighbors for physical models of mesoatom association during crystal formation.

Figs. 5 and 6 show the corresponding analysis of the local mesoatom packing for DD and DP, respectively. Relative to the DG mesoatoms, these structures exhibit a few key differences. First, the mesoatomic units of DD and DP are *achiral* and hence equivalent (up to rotations) between the two single network domains. Second, they exhibit a smaller fraction (less than half) of their contact with intra-network (a.k.a.

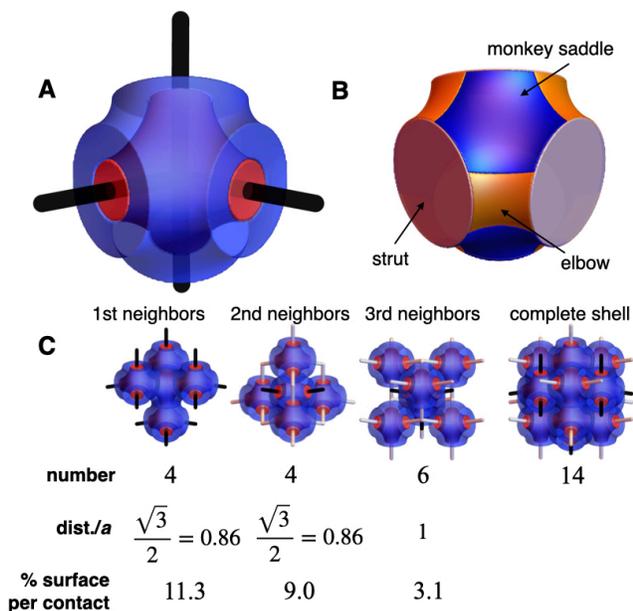

**Figure 5 –** Shape and packing of DD mesoatoms. (A) shows one of the elementary DD mesoatoms which is recolored in (B) in terms of the regions of contact between the 14 neighboring mesoatoms that touch its surface. These neighbors fall into 3 distinct sets according to their center-to-center distance and surface contact as detailed in (C).

"strut" neighbors): 45% for DD and 32% for DP. The DD packing is still similar to DG in that its closest (and highest contact) neighbors belong to the same network, with its second- and third-nearest neighbors belonging to the second diamond network domain. However, DP packing has a distinct pattern where its closest, highest-contact neighbors are its 8 monkey-saddle faces with the second primitive network domain, while its 6 lower-contact, strut neighbors are more distant yet belong to the same primitive network domain.

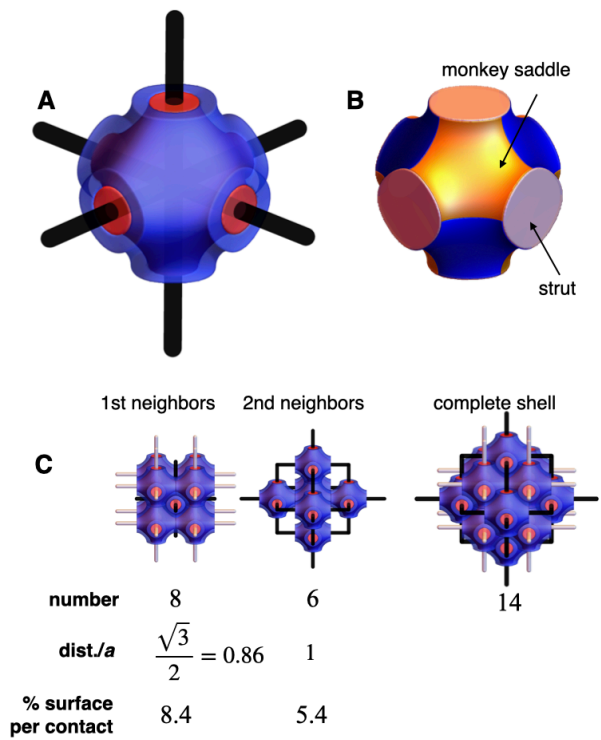

**Figure 6 –** Shape and packing of DP mesoatoms. (A) shows one of the elementary DP mesoatoms which is recolored in (B) in terms of the regions of contact between the 14 neighboring mesoatoms that touch its surface. Center-to-center distance and surface contact of two distinct sets of neighbors detailed in (C).

Taken as sequence, the mesoatomic shapes from DG to DD to DP represent a progressive increase in the number of point symmetry elements (and asymmetric motif copies) from 6 to 24 to 48, respectively, in addition to the corresponding increase in network valence. Along with this increasing symmetry, we observe a transition in the distribution of intra- and inter-network neighbors in the constitutive mesoatomic units of the double networks. In lower-symmetry/coordination structures, like-network mesoatoms are relatively closer and have higher contact, sheathed by a larger number of more distant neighbors of the opposing network. While in higher-symmetry/coordination structures, inter-network saddle contacts are pulled closer and strut contacts are pushed out, lowering the intra-network contact per mesoatom. Notably, for the highest symmetry DP mesoatoms, a distinct set of "elbow" contacts are absent, with the entire saddle surface taken up by a single set of "monkey-saddle" contacts.

Lastly, we note that the distinct neighbor correlations of mesoatoms also encode the local topology of the network assemblies, shown in Figure 7. For each of the cubic double networks, the inter-network neighbors in the first complete shell of neighbors (i.e. the contacting mesoatoms) compose the *f* elementary loops that catenate the *f* struts emerging from a central mesoatom (i.e. each of the struts emerging from a node is looped by neighbors in the contacting shell of mesoatomic neighbors). For DG, loops are composed of two series of 4 elbow neighbors that join at the two monkey saddle neighbors that sandwich the central mesoatom (i.e. loops of 10 total). As a set, the 14 inter-network mesoatoms form a "trihedral cage" that enmeshes the central particle, and encircle each of the like-network struts that emerge from it. Similarly, the 12 and 8 inter-network mesoatoms for DD and DP, respectively, compose tetrahedral and cubic cages for these networks.[2]

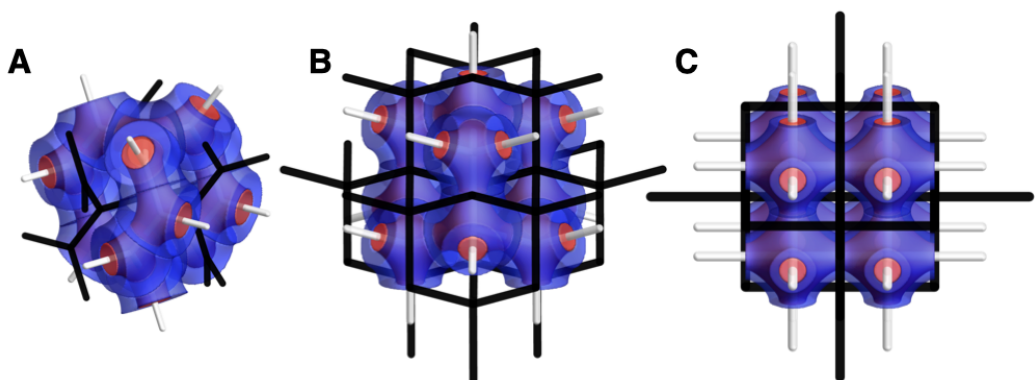

**Figure 7** – The set of catenating mesoatom neighbors on the "-" (network) that envelope a central mesoatom on the "+" network (black) are shown for DG (A), DD (B) and DP (C). The (black) skeletal bonds of the catenated "+" connected to the central mesoatom are shown to illustrate the local bond topology.

In the following section, we consider how the local geometry neighbor contact encoded in the shapes of mesoatoms plausibly templates the kinetics of inter-catenation in a simplistic model of assembly.

**IV. Mesoatomic implications for network crystal formation: a minimal model**

In the prior section we described the elementary mesoatomic molecular groups of tubular network crystals extracted using symmetry-based principles from the final equilibrium morphologies of neat diblock copolymer melts. In this sense, these shapes describe the "mature" state of these mesoatomic elements, as opposed to transient structures occurring as the system evolves towards the idealized equilibrium state. Here, we proceed one step further, to consider how the anisotropic shapes of these mesoatomic elements might provide a plausible basis to template the non-equilibrium kinetic pathways to long-range ordered and topologically non-trivial double network crystals. Noting that these so-defined mesoatoms pack perfectly to tile space, and moreover, possess outer terminal surface shapes that would likely promote and stabilize strong orientational correlations between neighbors based on shape complementarity alone, we ask the basic question, how would particles with the shapes of such mesoatoms assemble? In the context of block copolymer melt assembly, this model proposed makes the assumption that under relevant conditions, mesoatoms form first, and to good approximation, are driven to adopt the non-convex and high-symmetry shapes of their ultimate, mature forms. We discuss likely conditions where these assumptions may or may not be met below. Assuming preformed mesoatoms, we consider the simplest possible assembly-model for the local interactions between those mesoatomic units, and from this, model the non-equilibrium process of crystal growth and inter-catenation between constituent network domains. We leave the numerous open questions raised by this proposition (e.g. how might more realistic models incorporate dynamic evolution or cooperative distortion of mesoatoms into the assembly process?) for a later discussion.

---

[2] Wells describes these shortest network loops as 10-3 net, 6-4 net and 4-6 net [36].

*Model*: Our model considers simple non-equilibrium growth kinetics of TPN crystals via the sequential binding of nodal "particles" whose shapes and contact geometries are determined by the mesoatomic shapes described above in Figs. 5-7. We consider a process driven by a simplistic model of mesoatom interactions, in which the binding energy of cohesive interactions between mesoatoms is purely determined by their mutual surface area of contact. We consider in turn each of the DG, DD and DP structures: the DG with its 3 strut-bonds and 10 mesoatoms per loop, the DD with its 4 strut-bonds and 6 mesoatoms per loop and the DP with its 6 strut-bonds and 4 mesoatoms per loop.

Defining $\phi_{ij}$ as the fractional surface contact between neighbor mesoatoms *i* and *j* (i.e. $\phi_{ij}$ = 0.196, 0.121 and 0.014, for 1st, 2nd and 3rd neighbors of DG mesoatoms), the energy of a cluster of mesoatoms is

$$E \equiv -\sum_{<ij>} \phi_{ij}$$

where the sum is taken over occupied neighbor pairs of mesoatoms. Implicitly, this model neglects possible differences in the (free) energies of contact between strut neighbors (like-network) and saddle neighbors (inter-network), which could arise due to entropic differences of brush domains meeting parallel to or perpendicular to mean chain directions. Also, surface energy of different faces may vary due to enthalpic differences in the cohesive free energy density in tubular and matrix blocks and the composition differences at those faces, which e.g. would depend on the relative solvent quality and concentration for solution-cast assembly. For ordered phase formation from a higher temperature bulk melt state (via the ODT), there is obviously no effect of solvent on cohesive free energy density nor variation of the volume fraction of the subdomains via a preferential solvent. Neglecting these potential physical chemical factors implies that the differences in mesoatomic binding energies derive purely from the complementarity of their anisotropic shapes. In addition, we assume that rigid mesoatomic units only bind in perfectly oriented and spaced arrangements (i.e. their centers can only lie on the appropriate set of Wyckoff positions and adopt orientations consistent with the symmetry required by the space group of the ultimate crystal structure). Such a model considers mesoatom surface interactions to be extremely short range, and further assumes that the complementarity of the non-convex shapes restricts relative rotations. The latter effect is quite plausible for nested contacts between the 3-fold monkey-saddle faces, which provides a mechanism to template the long-range order and complex topologies of double network crystals even at the level of two-body contacts.

Assembly is modeled as a kinetic process following a very simple non-equilibrium dynamics, which adds the *N*+1 mesoatom to an existing cluster of *N* mesoatoms at the available location with the lowest total binding energy. Hence, after each step of the assembly process, the unoccupied neighbor positions of all *N* particles are scored according to the total binding energy of adding the mesoatom at that position, and the next mesoatom is added to the boundary position with the strongest binding. If there are multiple locations with the same minimal binding energy, one of those degenerate sites is simply chosen for mesoatom placement. In this way, assembly is modeled via a "greedy kinetic pathway" that lowers the energy the by the largest possible amount at each step, following a process that is irreversible and largely deterministic (with the exception of degeneracy of strongest binding positions). While there obviously is no guarantee that the state clusters are close to ground states, this simple algorithm allows us to explore the interplay between local packing, mesatomic binding and topological evolution via a plausible dynamic scenario for crystal growth. It is, of course, possible to consider more complex sampling approaches, for example, which attempt to consider near equilibrium assembly conditions at controlled chemical potential and temperature.

Below we first discuss the results of this simple model for growth of DG crystals, focusing on the evolving topology of double-network assembly. We follow this with a comparison to the growth of DD and DP crystals, and finally, briefly discuss results of the model for facet formation for growing crystals.

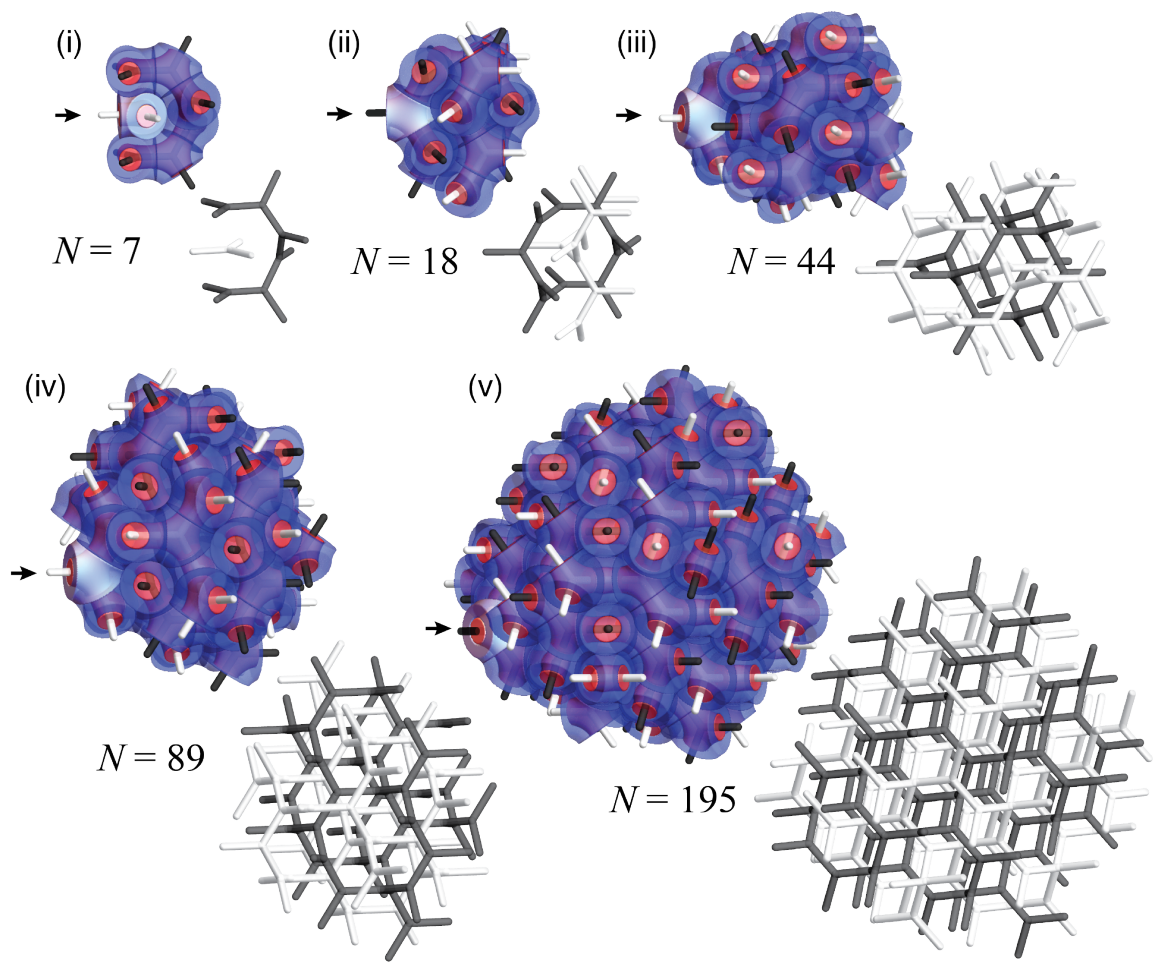

**Figure 8 –** Snapshots of the simulated growth of DG crystals for distinct numbers of added mesoatoms *N*. For each the most recently added *N*th mesoatom is highlighted by coloring the outer terminal surface of that mesoatom white. The skeletal bonds for occupied mesoatom positions are shown from the same viewing direction with the mesoatom volumes removed.

*Results - DG assembly:* We begin with the case of DG crystal assembly. Here, we note the kinetic growth algorithm considers addition of DG mesoatoms without explicit bias for chirality (i.e. both DG mesoatomic enantiomers maintain fixed, equal availability). Fig. 8 shows several snapshots of the growth of a crystalline DG cluster up to *N* = 195 mesoatoms showing both the space-filling structure of assembled mesoatoms as well as the skeletal bonds corresponding to those assembled mesoatoms. In each of those snapshots, the final ($N^{th}$) mesoatom added to the cluster is highlighted with a white shading of its outer terminal surface. In **Supporting Movie 1**, we show an animation of the sequence of mesoatomic additions for a cluster growing up to *N* = 89 units.

To understand how the local packing of DG elements templates the dynamically inter-catenation of the double network crystal, we analyze the co-evolution of binding energetic and network topology with increasing cluster size. In Fig. 9, we plot the binding energy per subunit, $E(N)/E$, as well as $\Delta E(N) = E(N) - E(N-1)$, the binding energy of the $N^{th}$ mesoatom, with points colored black/grey according to whether $N^{th}$ mesoatom joins the +/- single gyroid network, respectively.

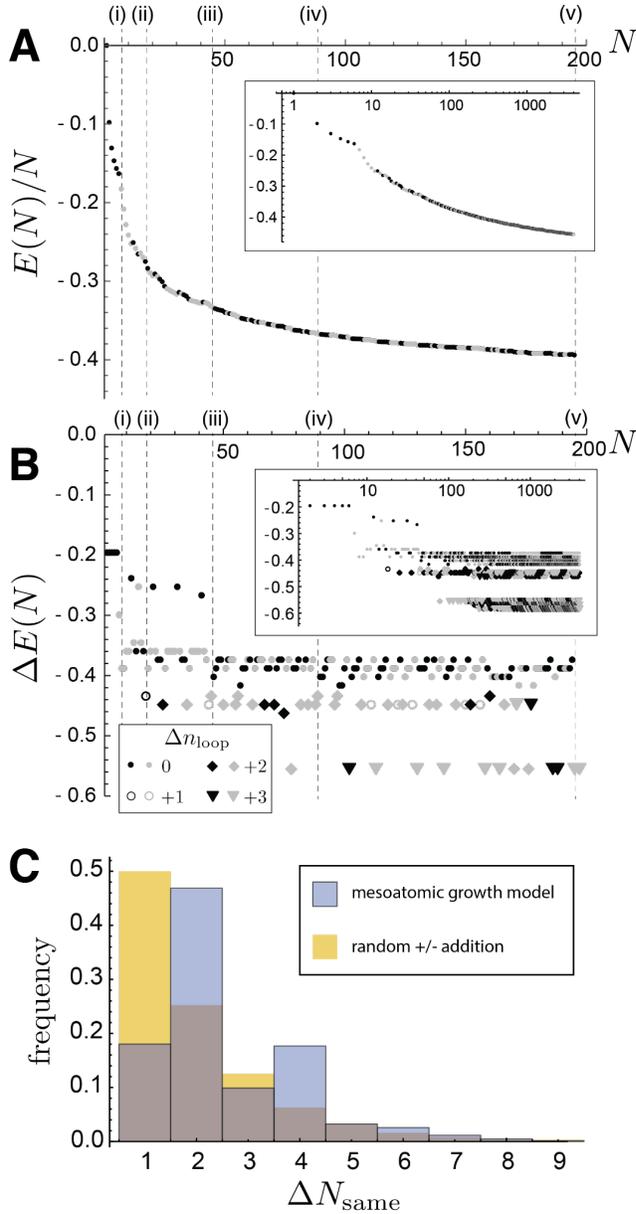

**Figure 9** – In (A) plots of the binding energy density of simulated DG crystals vs. mesoatom number, with each point colored according to "+" (black) or "-" (grey) network placement of the $N^{th}$ mesoatom. (B) plots the binding energy of the $N^{th}$ particle vs. $N$ and indicates the number of closed loops added upon binding. The dashed lines indicate snapshots (i-v) shown in Fig. 8. (C) Plots the frequency of consecutive mesoatom additions to the same networks of length, $\Delta N_{same}$, with mesoatomic simulations shown in blue and random (uncorrelated) additions shown in orange for comparison.

First, we observe a notable alternation for mesoatom addition between the networks, in which sequences of a few mesoatoms add to the same network before switching to the opposing network. The basic origin of this effect is easiest to understand for the first sequence of 7 mesoatoms (up to the first snapshot in Fig. 8.i). Because strut-strut face bonds are most favorable among all contacts, the first neighbor in the cluster is added as the closest neighbor in the same network. Such strut bonds within the same network remain preferred for the next several steps up to the $N = 6$ mesoatom because the total amount of saddle surface contacts a possible neighbor on the alternate network results in weaker binding than the $\Delta E_{strut} = -0.196$. This situation persists until the cluster forms 6/10ths of a closed loop of a single network. Such a structure is formed by black network mesoatoms in Fig. 8.i. For this configuration, it is straightforward to see that the partial loop envelops a high binding energy pocket, wherein a single additional particle (on the opposite network) can simultaneously form two monkey saddle bonds plus four elbow bonds resulting in a binding energy $-0.298 > \Delta E_{strut}$ which is stronger than an additional intra-network strut-bond. Hence, before closing the first loop, the assembly switches to the alternate network and proceeds to add to that network the next sequence of 5 mesoatoms. From this point, the mesoatom addition alternates back and forth, as partially completed network loops create new strong-binding pockets via their saddle faces.

Given this alternation of added particles between the networks, the formation of the first loop in the structure does not occur until the 18th mesoatom is added (snapshot of Fig. 8.ii), far in excess of the minimum 10 mesoatoms needed to form a single loop of the gyroid network. In Figure 9B, the binding events are labeled according to the number of network loops (0, +1, +2 or +3) added. Added loops are labeled as open circles (here "loops" are counted as the filled 10-mesoatom fundamental cycle of the gyroid graph), showing clearly that these binding steps are particularly strong binding events due to the addition of at least two intra-network strut-bonds, generically exhibiting lower energies than binding events that leave the topology unchanged. This suggests that cluster states that correspond to loop closure in the $N^{th}$ mesoatom addition will be particularly stable and relatively longer-lived states of growing DG crystals.

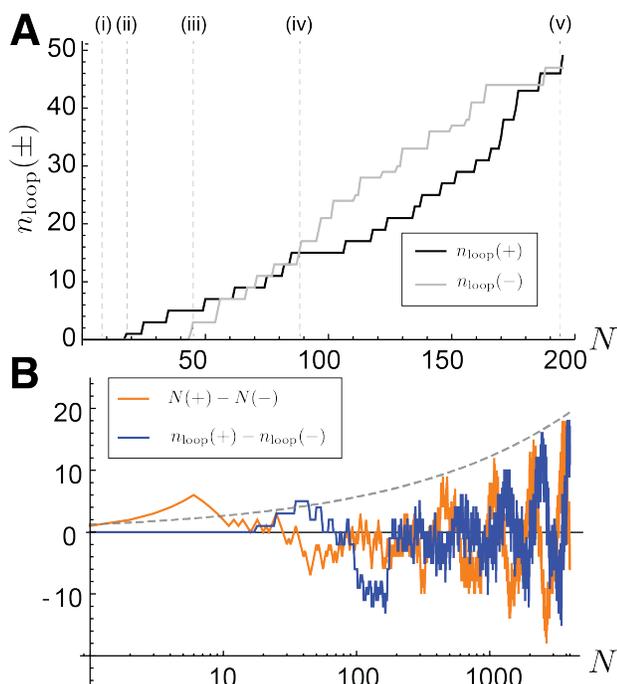

**Figure 10 –** In (A) plots of the number of closed loops in the + and 0 networks in the simulated DG crystal vs. mesoatom number. (B) plots the +/– excess of mesoatom number and loop number vs. number of added mesoatoms in the DG crystal. The dashed line has the scaling N^(1/3), discussed in the text.

In the sequence of the first 4000 mesoatom additions, 61% of the binding events leave the number of closed loops unchanged with the remaining fraction of corresponding binding events ~~that~~ increasing the number of loops by +1 (3%), +2 (11%) or +3 (25%). We note from Fig. 9B, that the most energetically favorable binding events tend add multiple loops (i.e. +2 and +3), consistent with the addition of a DG mesoatom along a <111> neighbor, creating at least two strut (intra-network) and one monkey-saddle (inter-network) bonds, and hence resulting in a large binding energy = -0.513.

We next analyze the alternating network growth kinetics in the model of DG growth in terms of the number sequenced mesoatoms that add to the same network ($\Delta N_{same}$). In Fig. 9C we plot the frequency of $\Delta N_{same}$ like-network additions for clusters up to $N$ = 4000. For comparison, we also plot the expected (exponential) distribution that would be expected if the subsequent binding to + or - networks was completely uncorrelated. This shows the relative excess of 2- and 4-mesoatom runs to the same network. While this is indicative that strong intra-network binding promotes like-network correlations, the $\langle \Delta N_{same} \rangle = 2.6$, which is much less than the length of 10-atom loop, consistent with the observation that additions switch back and forth multiple times between loop closure events in the DG crystals.

In Figure. 10 we plot the kinetics of loop formation in the growing DG cluster. We first focus, in Fig. 10A, on the number of loops $n_{\text{loop}}(\pm)$ in the two single gyroids (denoted as + or -) in the early stages of cluster growth illustrated in the highlighted snapshots of Fig. 8. Beyond the latency of the first loop forming after the 18$^{\text{th}}$ particle, we observe a surprising asynchrony in the looping of the two networks. The first three loops form in the same + network, well before the opposing - network forms even its first loop at the *N* = 44 mesoatom (shown in the snapshot in Fig. 8.iii). Following this, a rapid sequence of 2 loop additions in the - network quickly equalizes with + network, eventually overtaking looping in that network after the *N* = 89 mesoatom (shown in the snapshot in Fig. 8.iv). The cluster maintains an excess of - loops over a fairly large span, up to the *N* = 195 mesoatom (shown in the snapshot in Fig. 8.v), after which point the looping in networks remain fairly equal from several additional mesoatoms.

In Fig. 10B, we plot the differences in the looping between the two networks, as well as the difference between the total number of mesoatoms in each network, for cluster growth up to *N* = 4000. This shows that the initial loop imbalance roughly equalizes between *N* = 195 and *N* ~ 1000, but at longer times starts to exhibit more regular "sawtooth" pattern alternating swings of + or - loop excess. For *N* > 1000, loop excesses seem to be in lockstep with broader swings in the excess numbers of mesoatoms added to + vs. - networks albeit with a lag between loops and mesoatom excess. This, in combination with the fact their magnitude grows with *N*, seems to suggest these fluctuations are dictated by fairly regular patterns of surface layer growth, presumably with "steps" of the surface growth exposing different numbers of strong binding pockets on the like vs. unlike gyroid networks along distinct regions (i.e. facets) of the growing crystal. We return to the geometry of growing facets in the crystal below.

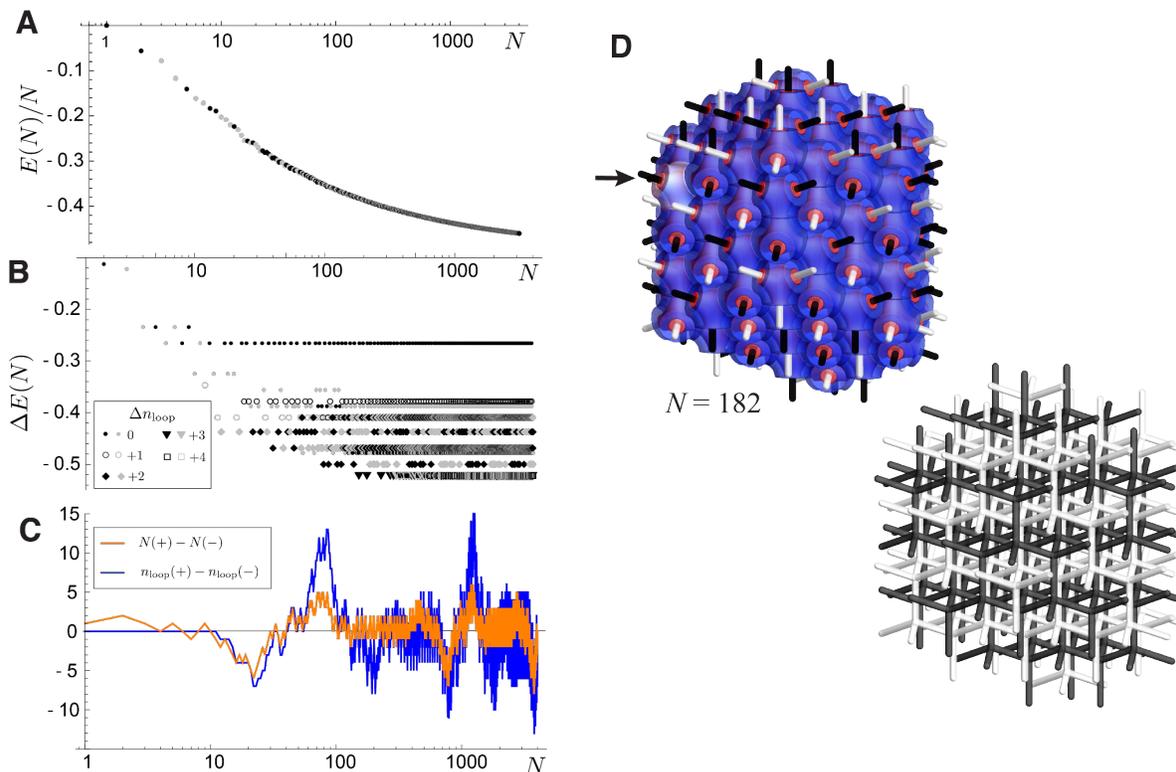

**Figure 11 –** Results of simulated mesoatomic growth DD crystals, showing energy density (A), binding energy (B), +/- network excess (C) vs. number of added mesoatoms for up to $N$ = 4000. (D) shows a snapshot of a growing cluster of simulated DD mesostoms, and the skeletal network of bonds corresponding to occupied mesoatom positions.

*DD and DP assembly:* The above results for the mesoatomic model DG assembly illustrate how the local shape and packing of the non-convex building blocks of double network crystals template the dynamics of intercatenation. Here, we compare results for our deterministic model of mesoatomic crystal growth for the higher coordination DD and DP networks.

As shown, in Fig 11 A and B, compared to predictions for DG (Fig. 9 A-B), addition of mesoatoms in DD alternate much more frequently between the disjoint networks, even at the early stages. For example, only the first 2 mesoatoms bind to the same network before switching to the lower energy binding on the opposite network for the third and fourth particles. This higher alternation reflects the fact that in comparison to DG, next nearest neighbor (inter-network) contacts are closer in surface area to nearest neighbor (intra-network) contacts for DD. The higher alternation is also consistent with the smaller loop size: 6 for DD compared to 10 for DG. Distinct from DG, as highlighted in Fig. 11B, DD assembly exhibits binding events that add up to +4 loops. Compared to simulated DG assembly, such events are likely enabled by the higher coordination (4) of the DD network. Indeed, the lowest energy binding events are triple or quadruple-looping events (such as the $N$ = 182 particle addition highlighted in Fig. 11D), corresponding to addition along a 3-fold <111> direction, forming three intra-network bonds, one "monkey-saddle" inter-network (2$^{nd}$ nearest neighbor) and three "elbow" inter-network bonds (3$^{rd}$ nearest neighbor). Within the first 4000 mesoatoms added, 35% binding events do not increase the number of loops with the remaining fraction adding +1 (5%), +2 (40%), +3 (7%) or +4 (13%) network loops. In Fig.

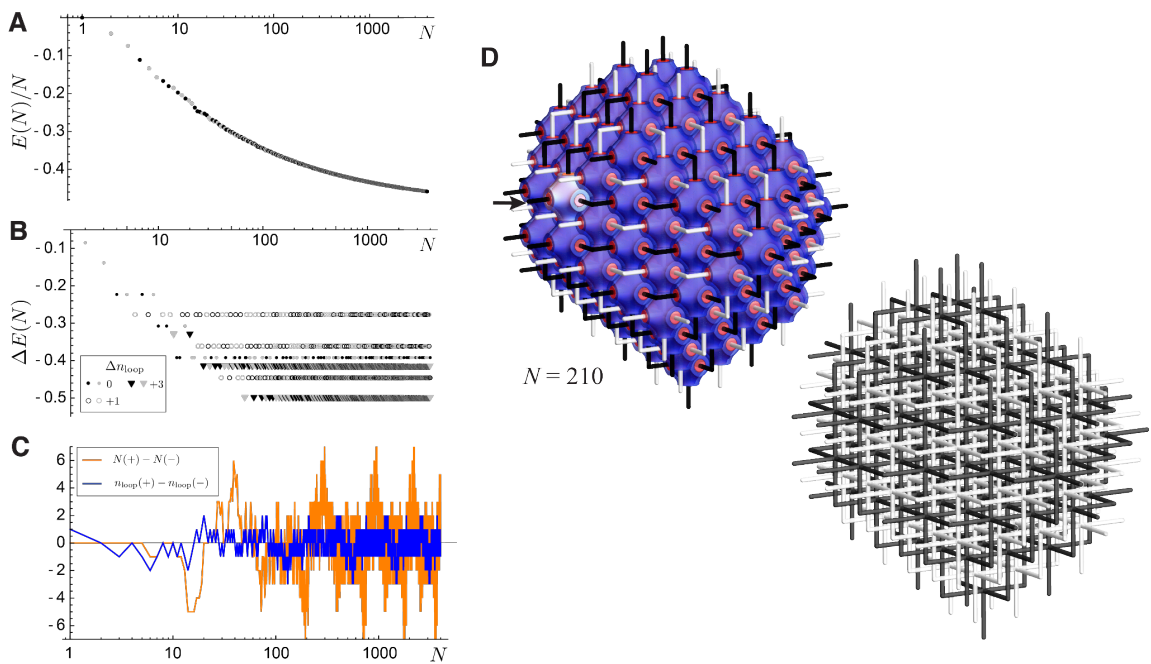

**Figure 12 –** Results of simulated mesoatomic growth DP crystals, showing energy density (A), binding energy (B) and +/- network excess (C) vs. number of added mesoatoms for up to $N$ = 4000. (D) shows a snapshot of a growing cluster of simulated DP mesostoms, and the skeletal network of bonds corresponding to occupied mesoatom positions.

11C we observe again the fluctuations in the addition to the two distinct networks of DD. This assembly also shows an initial period of + vs. - addition (and looping) imbalance at early stages, that recovers to a balanced crystal around $N \sim 100$ mesoatoms. However, unlike DG assembly, in DD crystals, fluctuations of network excess do not seem to show a coherent alternation, at least up for $N$ = 4000. Also, there is no significant lag between fluctuations of mesoatom addition to networks and the loop addition, with the latter tending to be simply proportional to the number excess of + vs. - mesatoms in the crystal.

Turning now to mesoatom assembly in the six strut-bond DP (results summarized in Figure. 12), whose mesoatoms possess the highest valence and smallest basic loop among the DG, DD, DP set, and who have stronger inter-network bonds than intra-network bonds, we find a several notable distinctions. First, the rate of alternation between network additions between + and - networks is the highest among the 3 cubic network crystals. The mean span of "like network" additions in simulated assembly for $N$ up to 4000, $\langle \Delta N_{same} \rangle$ for DP is only 1.15 compared to 1.7 for DD and 2.6 for DG. The enhanced tendency of DP to rapidly switch between + and - network additions (i.e. 85% of like-network spans in DP assembly include only a single mesoatom) is clearly a result of the stronger binding to inter-network (monkey saddle) faces than for intra-network (strut) faces in DP relatively to the other networks. Second, as shown in Fig. 12B, DP assembly is characterized by a surprisingly limited distribution of loop additions upon binding. For $N$ up to 4000, 97% of added mesoatoms increase the number of loops in structure and these are confined to only +1 (24%) and +3 (73%) loop additions. The especially "loopy" assembly is consistent with the fact that in the final DP crystal the ratio of loops to mesoatoms is 3:1, which is a much higher density of loops relative to DD (2:1) and DG (3:2). Last, we observe the that, like the DG assembly, the fluctuations in mesoatom and loop addition to + vs - network in DP crystals (Fig. 12C) falls into a regular alternating sequence after an early period ($N > \sim 100$), which we again attribute to a repeating pattern of surface growth with successive "layers" of crystal growth. However, unlike DG assembly (in Fig. 10B) the magnitude of these excess fluctuations does not appear to grow with $N$, suggesting that the network excess is associated with features of the crystal surface that *do not* grow with size (i.e. vertices of a faceted shape). In the following section, we return to this observation in the context of the emergent external shapes (crystal habits) of single crystals during simulation of double network crystal growth.

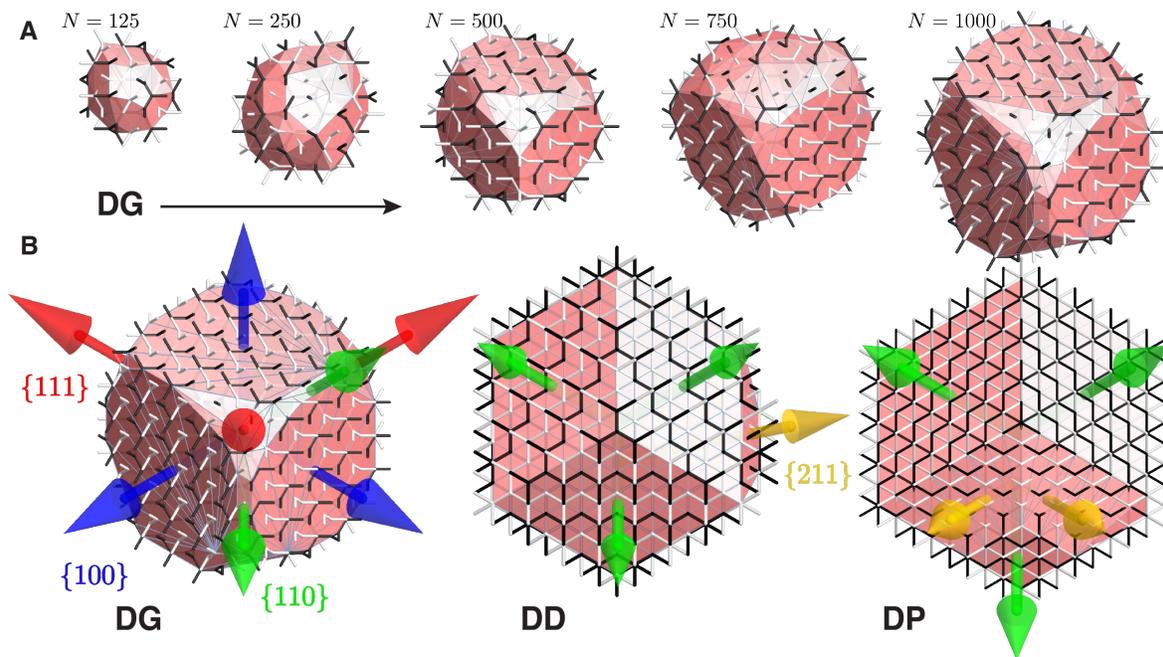

**Figure 13 –** In (A) growing clusters of DG crystals, with semi-transparent pink surface facets surrounding the centers of occupied mesoatom positions. (B) shows the crystal habits of simulated DG (left) DD (middle) and DP (right) at the N = 1500 snapshot, all viewed from a common <111> direction. The colored arrows highlight the orientation of prominent surface directions on the apparently faceted shapes of the crystals. The apparent breaking of the cubic symmetry in the distribution of facets is a result of the non-equilibrium growth pathways from which the snapshots are taken.

*Crystal habits:* Beyond a detailed picture for evolution of topology in intercatenated network crystals, the mesoatomic assembly model provides direct predictions of the external shape of growing crystals. Equilibrium crystal habits are generically described through the Wulff shapes, which derive from the anisotropic surface energies of distinct crystal facets. The local contact model described above is sufficient to fully determine the surface energetics of DG, DD and DP crystals (i.e. the distinct surface energies among various Miller planes). Again, while our deterministic kinetics are not guaranteed to sample ground state clusters for a given $N$, the model obviously favors growth on high surface energy faces (i.e. particular strong binding directions). Indeed, for sufficiently large clusters, we observe the clear formation of well-defined and stable faceting.

In Fig. 13A we show the external surfaces of an evolving DG cluster (viewed from the <111> direction), with the protruding skeletal networks.  The sequence shows that for relatively small clusters (e.g. *N* = 125 and 250) the cluster boundary appears roughly spheroidal.  But, by *N* = 500 and beyond, the surface shape of the DG begins to exhibit a characteristic pattern, ultimately growing into a cuboidal-shape with rounded corners and edges for *N* = 1000.  Subsequent snapshots of surface shape show fluctuations around this basic shape, but with the same dominant {100} faces showing apparently the same characteristic fraction of surface area at late stages.

Similar faceting behavior is found for DD and DP crystals, but with large clusters exhibiting different crystal habits.   Fig. 13B shows the *N* = 1500 snapshots for DG, DD and DP (all viewed from a common <111> direction), with Miller indices of the largest area facets labeled.  Notably, the largest faces of the DG crystal are the {100} planes, while in DD and DP, the facet planes are the {110}  (with minor facets along {211} for DD and DP) yielding habits roughly corresponding to rhombic dodecahedra.

Focusing on the crystal habits of the DG, in addition to {100} type facets, we also observe prominent {111} facets, leading to a somewhat rounded-cube shape. As well, there are smaller {110} regions along the edges.  We note that that these {110} facets have normals that correspond to directions of intra-network bonds, which are relatively strong binding energy compared to the inter-network "monkey" saddle bonds along <111>, and hence might be expected to possess relative high surface energies, and low facet areas in the corresponding Wulff shape.  Notably, the bonds that protrude through the {110} edges for the cuboidal shapes appear to be dominantly of one network chirality (i.e. - in the case shown in Fig. 13 B).  Thus, if the + vs. - excess derives predominantly from these edge-regions of the cuboidal crystals, we would expect the fluctuating chiral excess to grow with the edge length, as $N^{1/3}$ .  This scaling is consistent with increasing magnitudes of mesatom excess for DG assembly in Fig. 10 B (dashed line).  This suggests that directional energetics of mesoatom binding could give rise to spontaneous fluctuations of surface chirality (i.e. + vs - excess) that grow arbitrarily large with crystal size.  In contrast with the growing "asymmetry" of DG crystals with *N*,  the + vs - network excess of DP appears to be constant with *N*, suggesting that this excess is associated with the *vertices* of the quasi-polyhedral crystal (for rhombic dodecahedra these correspond to the eight <111> and six <100> directions).

**Discussion**

A generic construction of the elementary mesoatomic units of supramolecular network crystals, focusing on the cubic double networks of diblock copolymer melts was proposed with the DG as an illustrative detailed example.  This generalizes the notion of micellar groupings of molecules that constitute building blocks of 3D crystalline or 2D columnar arrays of sphere- and cylinder-like domains, respectively, which are ultimately confined to quasi-polyhedral volumes that tile the given crystal.  Like those cases, mesoatoms of network crystals are associated with maximal-symmetry subvolumes of domains within the equilibrium network crystal (i.e. the set of Wyckoff positions within single domains with the highest point symmetry).  Unlike spherical or cylindrical domains, however, mesoatoms in double-network crystals are non-convex shapes and derive from two types of faces that divide nearest neighbors: planar faces separating like-network neighbors and saddle-shaped faces separating adjacent neighbors on the opposing network.

The mesoatomic construction of network crystals provides a useful structural description of supramolecular network crystals, breaking their complex structure into local motifs, akin to more familiar cellular (e.g. Voronoi) constructions for compact domains. Going beyond this purely descriptive notion, we conjecture that this symmetry- and geometry-based deconstruction provides physical insight into collective properties of network crystals and plausible kinetic pathways by which they form. The three various inter-cantenated tubular network structures ultimately stem from constraints of packing non-convex mesoatomic shapes as well as expected differences in physical contact between domains along distinct faces: e.g. saddle faces sit at contact between opposing (matrix) domains, whereas strut faces include contact between both minor and matrix components. The latter distinction suggests an analogy between mesoatomic crystals and atomic crystals in which we view faces that divide brushes on opposing network domains as analogues of non-covalent (i.e. van der Waals) binding, while we associate faces composed of contact between multiple components (strut-bond faces) as analogs of covalent binding. In this analogy, the total cohesive energy between mesoatoms in double-network crystals includes both "covalent" and "non-covalent" contributions, whereas in crystals of sphere-like domains, where shapes are polyhedral relatives of Voronoi cells, inter-mesoatom binding is purely of the non-covalent type.

In the model introduced here we considered the distinct physical effects of various types of contacting faces to derive the assembled structure purely from the amount and type of same-block to same-block surface area of contact, implicitly assuming that free energy of surface contact is independent of which components are in contact, and how chains are oriented, across the contact faces. It is straightforward to consider generalizations of this simple binding model that relax this constraint. For example, considering the growth of double-network crystals in a solvent that is selective for one or the other component, it is possible to consider how the relative strength of binding along distinct faces would change depending on the both the area fractions of minority and matrix components along each mesoatom face, as well as the relative surface energetics of solvent contact to those components. For the situation where mesoatoms are forming and then assembling into crystals in a selective solvent for the matrix, there would be a correspondingly higher binding on along strut bonds as these shield the minority domains from solvent contact. This in turn would impact predictions for binding and catenation dynamics, as well as for the facet formation in large scale-crystal structures. In this context the mesoatomic model provides a natural and predictive framework to understand how highly-interconnected topologies for supramolecular network crystals form, based on local rules based on packing and binding thermodynamics.

The essential elements of the mesoatomic growth model described here are predicated on the following propositions:

1) The dominant pathway for network crystal formation is one where micelle-like groups (mesoatoms) of chains break isotropic (i.e. spherical) symmetry into lower-point group symmetry, with non-convex shapes *before assembly with other mesoatoms into crystals*
2) The optimized packing of the non-convex, asymmetric mesoatoms derives from both covalent-like and van der Waals-like binding which dictates the ultimate crystal space group symmetry and topology of the crystalline assembly
3) To a good approximation, the shape and packing characteristics of the "primordial" mesoatom can be derived from the structure of its ultimate, mature shape in the final crystalline state observed experimentally and computed theoretically.

Each of these propositions raises open questions for experimental and theoretical studies of actual supramolecular network assembly. In our analysis above, we restricted our focus to diblock copolymers, but the relevance of mesoatoms clearly extends to other macromolecular contexts where these or similar morphologies occur. We offer some brief comments about extensions of mesoatom concepts to other molecular architectures below.

The proposition, that non-convex mesoatoms form first, in the assembly process, may be reasonable on its face, but raises several important questions: do the thermodynamic prerogatives of molecular groups, due to the balance of entropy and enthalpy *within those groupings alone*, select the complex non-convex shapes, compatible with the nodal-interconnections of the ultimate networks with their characteristic inhomogeneous local thickness and (negative) curvature IMDS shapes? If so, this suggests that it should be possible to identify some range of thermodynamic conditions where individual (isotropic) spherical micellar domains (near to, but slightly above the critical aggregation conditions) break symmetry into the elementary trihedral, tetrahedral and octahedral symmetries consistent with the "mature" mesoatoms of DG, DD and DP, respectively. This raises a further condition about how different the shape and symmetry characteristics of the "primordial" mesoatomic elements might be from the ultimate "mature" bonded mesoatoms in the crystal. For example, the schematic in Fig. 14 shows a hypothetical symmetry-breaking pathway from a spherical micellar grouping into a trihedral micelle, a primordial version of the DG mesoatom. For this case, we imagine that the degree of warping of the trihedral unit could be quite variable. For example, the lengths of the "strut" like regions may be somewhat different from the geometry compatible with the final DG crystal packing. Additionally, depending at least somewhat on solvent selectivity, the degree of chain relaxation at the ends of the strut-like poles of the mesoatom will which modify the orientation and relative exposure of distinct chain portions along what ultimately become "endcaps". The shapes and thermodynamics of these endcaps would, in turn, have influence on the kinetic and thermodynamics of intra-network (i.e. strut) bonding in the subsequent DG crystal formation.

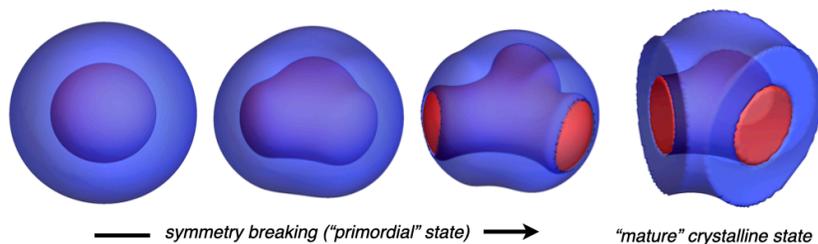

**Figure 14** – Schematic illustration of hypothetical symmetry-breaking pathway from isotropic spherical domains to trihedral $D_{3h}$ symmetries to of "primordial" $D_3$ DG mesoatoms (Wyckoff 16b).

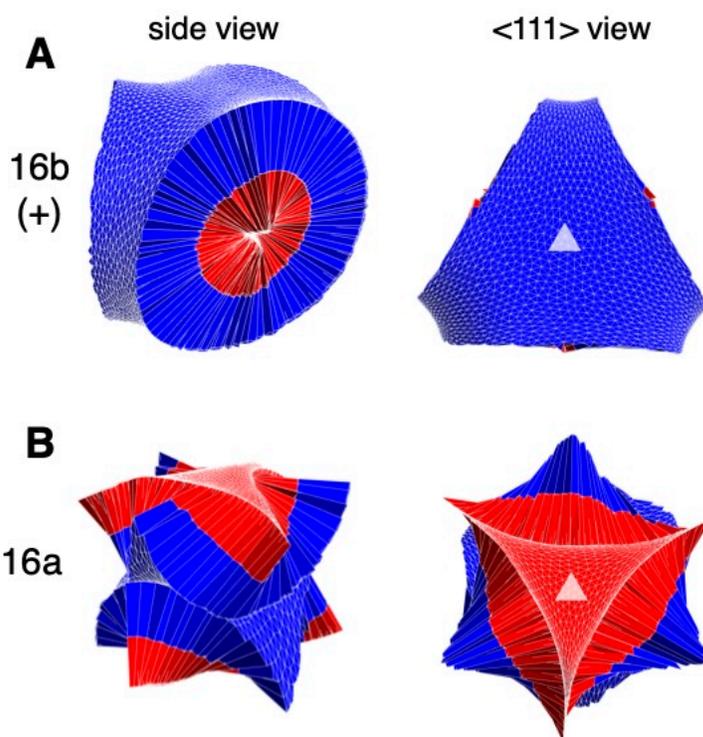

**Figure 15 –** Comparison of expected shapes of 16b mesoatoms (A) to putative 16a mesoatoms (B) of based on the medial tessellations of DG. Triangular prismatic regions model mean chain trajectories extending between terminal boundaries. While 16b mesoatoms are bound only by the TMPS-like (outer) terminal surface of the matrix block, large fractions of 16a mesoatom surfaces are bound by web-like (inner) terminal surface of the tubular block. In <111> view, only the 3-fold axis is highlighted of the respective $D_3$ (16b) and $C_{3i}$ (16a) are shown.

We note that our primary heuristic for identifying which particular positions of the ultimate crystal correspond to kinetically favorable groupings is purely based on topological and symmetry grounds (i.e. sub-regions of domains with maximal point symmetry). There are likely many conditions where double networks with a kinetic bias for other groupings. For example, should thermodynamic conditions at which mesoatoms first aggregate favor surface contact of one domain over another (i.e. non-selective solvent), the addition differences in the relative surface exposures of A or B-type domains could bias assembly towards other high-symmetry points. In Figure 15, we show comparative renderings of the mesoatoms of DG based on both 16b and 16a Wyckoff positions of $Ia\bar{3}d$. These constructions exploit a more refined strong-segregation packing description [30] which includes distributions of chain trajectories modeled by prismatic wedges extracted from medial surfaces of gyroid surfaces that model the terminal boundaries in the matrix and tubular domains. In this case, we observe that the strut faces of the 16b mesoatom must be at least slight non-planar to avoid cutting chain trajectories. The volume of the 16a mesoatom, is otherwise, markedly different, in that its surface is not bound by the TPMS-like terminal boundary of the matrix domains, but instead the twisted-web shape of the terminal surface of the inner domains, which closely approximates the inner medial surface of the IMDS. As a consequence of more complex and disconnected geometry of these bounding terminal surfaces, the putative 16a mesoatom is clearly more complex in shape, with an even larger surface to interior volume ratio than the 16b position. Additionally, because it includes subdomain regions from each of the two gyroid networks, it possesses two disjoint regions of the IMDS, as opposed to the single IMDS patch of the 16b mesoatom. Presuming that thermodynamics of IMDS formation is dominant in the formation of primodial mesoatoms, this suggests that the 16a mesoatom type would require two IMDS nucleation of events to form, as opposed to the single IMDS nucleation for the 16b mesoatom, and hence kinetics of mesoatom formation of the 16a mesoatom (or any other mesoatom composed of fragments from multiple domains) would likely be much slower than the 16b.  However, it is possible that, under conditions where increased surface exposure of the minority domains is favorable over the majority domain, nucleation of 16a-type mesoatoms could preempt formation of the 16b-type mesoatoms.  Notably, unlike the 16b mesoatoms, 16a positions are achiral, and are not described by a tubular junction motif, but instead, a double-layer minimal saddle patch.  While the shape and local contact of such a distinct domain will template altogether distinct assembly kinetics, is it straightforward to consider how to extend the analysis and arguments presented here to these alterative shape and symmetry mesoatoms.

The second and third propositions, that packing the non-convex mesoatomic shapes templates the ultimate crystals formation, raises an important question about the *malleability* of mesoatoms.  Mesoatomic groupings are composed of large numbers of flexible molecules.  For example, in the DD and DG assemblies from a PS-PDMS diblock reported in refs. [24,37] one can calculate that the respective tetrahedral and trihedral nodal volumes possess roughly 2,500 and 1,100 chains (assumption of a single DD mesoatom giving rise to two DG mesoatoms would predict a ratio of 2:1).  Owing to their many mobile, flexible and independent constituent parts, mesoatoms are inherently malleable objects, and the thermodynamics of their inter-mesoatomic packing takes place at a similar free energy scale to thermodynamics of their internal rearrangements.  This means that mesoatomic shapes are in reality far from static, and likely evolve and adapt significantly during the binding event to a growing crystal.  As alluded to above, binding along intra-network struts by match-up of the respective block regions across the strut-bond is likely to require some degree of radially "combing" chains along the normal to the skeletal graph at the endcaps of the priomordial mesoatom (e.g. Fig. 15).  It is also reasonable to expect the shape of outer "saddle skin" of mesoatoms to adjust somewhat as opposing brushes come into close contact.  Notably, for DG mesoatoms, there is an additional question about when and how primordial mesoatoms that compose the alternate + and - networks break achiral symmetry.  One possibility (consistent with the assumptions of our minimal growth model) is that primordial mesoatoms themselves are unstable, spontaneously breaking symmetry into distinct populations of opposite chirality, and this pre-existing chirality organizes the subsequent kinetics of crystal formation.  An alternative scenario, arguably more plausible for achiral constituents, is that primoridal mesoatoms of DG are achiral and become chiral upon binding and adapting to the intercatenating DG crystals.  These effects all suggest a more realistic physical model of mesoatomic assembly will require malleability of the shapes and inter-mesoatom correlations, most crucially allowing bonding along different "faces" to take place over a more flexible range of angles and distances.

Several classes of discrete particle models have been developed in recent years that incorporate anisotropic binding directions and strengths, mimicking the key features of our mesoatomic particle model.  These include models of convex hard polyhedra [38], whose complex shapes and symmetries lead to the formation of rich array of crystalline and liquid crystalline morphologies, purely due entropy and close-packing considerations (i.e. what has been dubbed

"entropic bonding" [39]). Relative to such models, our description of mesoatom assembly assumes that binding is cohesive, and likely more important, that the nesting of *non-convex* particles is necessary to guide the formation of properly intercatenated double networks. Beyond such hard particle models, a range of "patchy sphere" modes have been explored in the recent decades motivated by questions as broad as colloidal glass formation, functional DNA liquid assembly and thermodynamic anomalies of water [40–42] . These typically involve building short-ranged "sticky" patches on otherwise isotropic (spherical) cores [43], with fixed number and symmetrical arrangements. Notably, these models parameterize a degree of angular fluctuation in the binding, which would serve as proxy of mesoatomic malleability [44]. However, at present, most such models include only attractions along "strut" directions of what might ultimately result as like-network contacts (e.g. only trivalent or tetravalent sticky bonds). In light of the above results, it would be interesting to understand how the incorporation of attractive interactions along directions that enable inter-network binding would influence the thermodynamics and kinetic accessibility of inter-catenated double-network crystal formation. Lastly, we note the existence of network forming models where the local building blocks (i.e. the mesoatoms) of the crystals themselves are composed of at least of few distinct particles. The interactions between the components of those mesoatomic motifs both template the stable local symmetries of those units, as well as their flexibiltiy and potential ability to reconfigure between different types of mesoatoms, much like what would be expected for supramolecular mesotoms. Based on a binary mixture of two classes of attractive particles, research by Molinero and coworkers has developed and explored a model in which DG crystals compete with lamellar and columnar phases, leading to rich insights into phase formation, transformation pathways, nucleation and growth of DG crystals [45–47]. A model of anisotropically sticky spheres developed by Chakrabarti and coworkers [48,49] has been shown to exhibit assembly into tetravalent and hexavalent network crystals. However, at present only single networks (e.g. single diamond [44]) have been observed in simulations of these models, presumably, because the close packing of bound spherical cores obstruct the incorporation of a second intercatenated network. Certainly generalization and extensions of such models as coarse-grained representations of mesoatoms hold potential for more extensive studies of equilibrium and non-equilibrium assembly from mesoatomic units. One challenge will be to parameterize the complex shapes, anisotropic binding *and* deformability of this coarse-grained mesoatoms in terms of physical models that connect to intra-domain deformation of supramolecular packing within those units.

The malleable nature of mesoatomic elements has important consequences not only for the fluctuations of local bonds/bond angles in growing crystals, but even more profound consequences for the complex possibilities of distinct states of disorder or defects in network assemblies, as well as pathways of interconversion from one type of network to another. Next we describe this and other experimental implications for the putative mesoatomic building blocks of network crystals.

**Experimental Observations: Searching for Mesoatoms**

The conjectured notion that mesoatomic units extracted from the "final" ideal double-network crystals are the natural "building blocks" of these structures leads to three key questions :

1) What is the current experimental evidence for the existence of mesoatoms?
2) Does this evidence allow determination of their symmetries and shapes?
3) For a given processing pathway, how do the (malleable) mesoatoms adapt during their assembly and longer-term maturation to and within an ordered assembly?

Answering these questions requires investigations that allow identification of the morphological characteristics of purported mesoatoms and their aggregated structures. Quantitative characterization of mesoatoms would require measurements of geometric features such as the strut directions, angles and lengths, mesoatom volume and surface area and the shape of the outer terminal surface and the inner IMDS, following the evolution of (perhaps) the isotropic point group (K) to the mature DG chiral mesoatom $D_3$ point group (e.g. Fig. 4 A), as well as the statistical and dynamical variability of these metrics. As suggested by our minimal model (e.g. Fig. 8), mesoatomic assembly implies rich geometrical and topological information on cluster growth, loop formation and inter-catenation (e.g. Fig. 8(i)) which requires additional characterization at the inter-mesoatomic scale. At the largest scale, as the cluster becomes a coherent crystal, details on the nuances of faceting of the cubodial crystal (e.g. Fig. 13 A) are needed.

Crudely speaking, one can expect two classes of kinetic pathways where mesoatomic elements form: 1) mesoatom-first formation followed by mesoatomic aggregation into ordered structures and 2) spinodal-like formation of randomly connected network-like domains, which later mature into ordered structures.

This first path ("mesoatoms-first") suggests conditions at early times of an initially disordered (mixed) system, evolving into discrete primordial, micelle-like groupings, *before* their association into multi-mesoatom arrangements of the type that ultimately become ordered crystals. Primordial mesoatoms would be expected to from an initially homogeneous melt by cooling or from a homogeneous solution by solvent evaporation, under conditions where local packing *within* micellar groups leads to symmetry breaking into non-convex shapes (e.g. Fig. 14). As the local chain packing motif within these tubular networks reflects a negatively-curved (saddle like) IMDS, it is natural to expect such trihedral (DG) or tetrahedral (DD) micelles to from when prerogatives of chain architecture, stiffness and enthalpy favor local shapes intermediate to cylindrical or layered morphologies, but otherwise find themselves in micelle-like groupings. Capturing their formation requires experimental techniques with excellent spatial and temporal resolution in order to follow the kinetic structural path. The processing routes that may give rise to mesoatom-first assembly are two fold. In a neat system (i.e. pure diblocks) near to the binodal curve (but outside of the spinodal region), it may be possible to image the formation of primordial network mesoatoms that take the form of non-convex micelles coexisting with disaggregated chains. Such is the natural picture for the disordered sphere phases that from at the high-$\chi N$ and high-compositional asymmetry regime of diblocks, but for formation of mesoatoms of a tubular cubic phase implies that chain compositions are likely closer to regimes favoring packing intermediate to lamellar or cylindrical morphologies, where the gap between binodal and spinodal regions is typically smaller. Therefore, casting the assembly from a volatile solvent, may be particularly important for conditions where mesoatom-first assembly takes place, since it may allow conditions where formation of primordial mesoatoms with complex shape is favored, but at initially dilute conditions. However, as a two-component system this processing route introduces the complexity that as solvent is evaporated, the appearance of a variety of mesoatoms and mesoatom aggregates would depend on the solvent concentration and solvent quality for each block, including the possibility for the formation of alternative mesoatoms leading to nonequilibrium, metastable phases. An advantage for the solvent processing approach is the ability to possibly create individual mesoatoms and to follow their subsequent interactions and assembly.

The second path ("sponge-first") to forming and observing primitive mesoatoms is arguably simpler, which would follow from cooling a single component diblock from its high temperature disordered near-homogeneous melt state to a temperature below the order-disorder transition temperature (ODT), where over time, compositional fluctuations may lead to (nearly) spinodal decomposition into a disordered microphase separated state (sometimes referred to as a "sponge" phase) which then evolves via nucleation and growth into the ordered crystal from the parent "disordered network" state. In this path, a stage where mesoatoms are observed as individual, disassembled units may not even exist, but nevertheless, we posit that the collective behavior of the system, most importantly, its longer-term evolution to an ordered structure is likely to be controlled by the collective reconfiguration of these local groupings. Detailed structural analysis of sponge phases is difficult but would be highly informative concerning the possibility that the sponge phase may be comprised of a network of primordial mesoatoms. Evolution of the sponge phase by up-quench experiments controlling the annealing temperature and time could induce mesoatom evolution leading to nucleation and growth of ordered clusters from within the sponge phase.

The two main experimental techniques for elucidation of structure are scattering (reciprocal space) and imaging (real space). Scattering patterns can be readily collected during the transition from the homogeneous state to the mesoatom state as the sample environment (temperature, solvent concentration) is systematically altered. The main issue is unambiguous interpretation of the scattering curve. Direct observation techniques can be much more specific but since the typical mesoatom feature sizes are on the ~ 1-10 nm length scale, observation on these length scales requires electron imaging. This approach is much more challenging due to the need for a vacuum environment around the sample (especially for samples containing volatile solvent) and the deleterious interaction of the beam with the organic specimen as a function of electron dose.

For both techniques, discrete mesoatoms would be first detected due to the emergence of electron density contrast from the surrounding near zero contrast homogeneous medium. However, since the mesoatoms that initially form will be spatially and orientationally uncorrelated, the scattering pattern, even from very small regions, will be a superposition of the isotropically averaged shape transform of the mesoatoms, greatly compromising the extraction of specific characteristic descriptors. While the xray experiments can conveniently follow the kinetics of mesoatom assembly, quantification of the morphological descriptors appropriate to the symmetries of mesoatoms by xray diffraction only becomes possible when the evolving structure exhibits coherent clusters of many tens to hundreds of unit cells when characteristic Bragg peaks appear and allow detailed analysis of the coherent crystalline state, likely limiting the ability to track evolution of mesoatomic shapes via time-resolved scattering approaches.

In principle, microscopy experiments can reveal much more specific detail, especially during the birth and assembly-growth of mesoatoms but are unable to follow kinetic evolution in real time in a specific region due to accumulated damage from the electron dose. Generally, a series of observations are taken on different regions made by systematic structural arrest by quenching at various stages of the transformation. Capturing and analyzing the individual mesoatoms and mesoatom aggregates that nucleate and grow during solvent removal is an open challenge but advances in cryo-TEM and cryo-SEM originally developed for aqueous biological systems and liquid crystalline materials are proving valuable for elucidating the emerging structures in polymer solutions. The liquid specimen (e.g. solvent plus surfactant) is first vitrified outside the instrument at some stage of the process, transferred and imaged *in situ* at low temperature. The sample, typically a surfactant/water solution, is prepared and concentration and/or temperature adjusted to create, for example, a micellar species. For cryo-TEM, a thin specimen is made via processing within a controlled environmental vitrification system allowing extremely rapidly quenching (~ $10^{10}$ K/s) of the thin (~ 100nm) sample into liquid ethane in order to vitrify the structures and fluid matrix [50,51]. A specialized cryo-transfer TEM holder maintains the specimen at low temperatures (~ 77 K) and allows transfer into a cold-stage TEM for imaging. Such experiments have revealed details of micelle formation and aggregation. It is interesting to note that surfactant/water systems often have an equilibrium microemulsion phase that is described as a disordered membrane network structure (denoted as L3 or the "sponge phase") [52,53] that can be imaged by vitrification-arrest experiments. Research on the solution behavior of amphiphilic macromolecules [50,51] has adopted the same techniques for successful imaging of mesoatoms of spherical and cylindrical micelles [54] as well as for nodal "Y" junctions between wormlike micelles [55].

Recently, variable temperature liquid cell (VTLC) sample holders for TEM have been developed with thin, electron-transparent windows, enabling direct, real-time imaging of nanoscale assembly, although again, with the need to respect electron beam specimen alteration with increasing dose [56]. Gianneschi et al. performed careful assessment of electron damage with dose and made VTLC TEM measurements of transient intermediate structures from thermally induced rearrangement of the inner core region of ABC triblock spherical micelles. Variable temperature SAXS showed changes in the form factor scattering from the core region in good correspondence with the TEM observations [57]. For BCP-solvent systems, future experiments would also require a means to uniformly decrease the solvent content of the sample during the observation in order to follow the kinetic pathways of aggregation. Thus, electron microscopic approaches are the best means to find and analyze primordial mesoatoms. Future application of VTLC and cryo-EM techniques for imaging the primordial mesoatoms in tubular network polymers seem very promising.

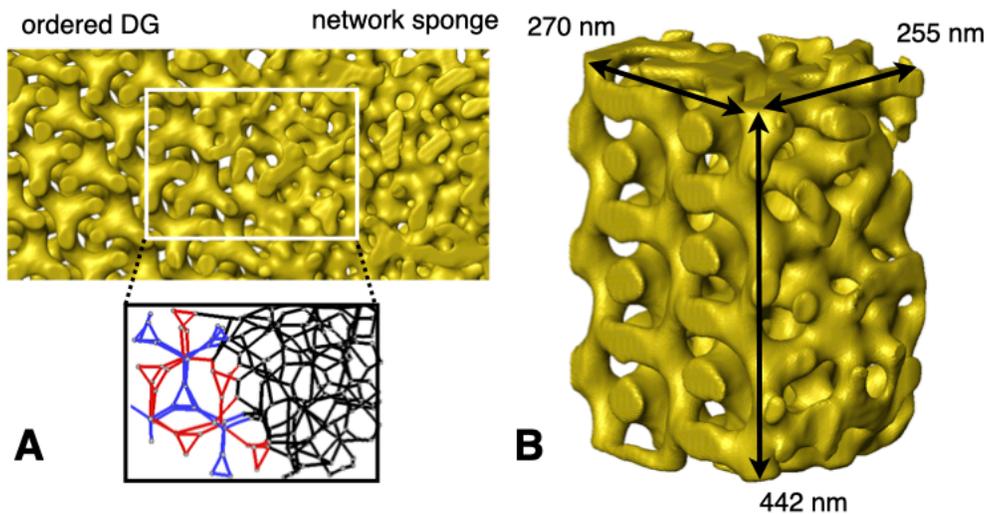

**Figure 16** 3D SVSEMT reconstruction of a region of a PS-PDMS diblock showing the boundary between a grain of the DG crystal with intercatenated trihedral networks (left) and a region of a sponge-like phase (right). (A) shows a skeletonization of the highlighted subregion spanning the boundary the interface between ordered and random networks. Skeletal graph bonds are shown as red/blue on the alternating gyroids of DG, while the random network bonds (a single fused network) are shown as black (B) Shows a volume with spatial dimension for scale. Over 88% of the units in the disordered network are trihedral with all mesoatom units belonging to a single network. (figures courtesy W. Shan). Note: This sample is not from spinodal decomposition, rather rapid evaporation of a solution (e.g. ref. [37]).

TEM images are 2D projections and while for simple geometries their 3D interpretation is reasonably straightforward, the disordered sponge phase as well as small regions of growing ordered network with their complex 3D inter-catenated structures requires tomographic reconstruction (electron microscopy tomography, EMT) for unambiguous structural interpretation. To date, there has been only one 3D reconstruction of the disordered network-like macromolecular sponge phase alongside ordered grains of DG [53] but for a diblock-homopolymer blend. The sponge phase was reconstructed for a volume containing about 70 unit cells and the bicontinuous network structure displayed a dominance of trihedral features (89% of the nodes of the associated skeletal graph had 3 functionality) along with some small perforated layer-like regions. Further analysis of regions where a growing DG cluster is evolving from an adjacent disordered region would shed much light on possible mechanisms for the disordered phase to ordered phase transformation and the role of malleable mesoatoms. A relatively new electron microscope technique, slice and view scanning electron microscopy tomography (SVSEMT) provides systematic imaging of much greater regions than TEM of thin sections – indeed, many hundreds to tens of thousands of cubic micron volumes (containing ~ 10 million unit cells) can be 3D reconstructed [37]. SVSEM involves creating a series of images at different depths of a sample by using a low voltage electron beam to image the near surface of the sample, followed by ion beam milling to remove a thin (~ 3 nm) slice of the sample, repeated over and over to produce a high fidelity 3D tomographic reconstruction with ~ 10 nm feature resolution (see Fig 16).

An interfacial region between the gyroid sponge phase and a DG grain is shown in Figure 16 for a PS-PDMS diblock (for details of sample and methods, see [37]). The viewing direction of the reconstruction can be chosen based on software manipulation of the 3D data and at the left is along the [111] direction of the DG grain. The inter-catenated minority component PDMS networks are readily identified, and their topology and geometry can be quantified by skeletonization of the 3D reconstruction. Interestingly, skeletonization of the adjoining sponge phase shows that ~ 90% of the nodes in a volume of $10^8$ nm$^3$ are trihedral units. Analysis of the strut directions in the sponge phase gives a nearly isotropic distribution, while, as expected, within the ordered grain, the struts are all well aligned along the <110> directions of the unit cell. Due to the orientational disorder, dihedral angles between adjacent nodes which are used to determine network chirality in the ordered DG phase, have a near isotropic distribution in the sponge phase. Moreover, in the region examined, there are no discrete, non-network PDMS regions, rather the minority PDMS component forms a *single* continuous network but without loop inter-catenation. This disordered network structure is not unlike the atomic scale continuous random network (CRN) model previously proposed for amorphous semiconductors (i.e. f = 3, amorphous arsenic would correspond to disordered DG, while f = 4, amorphous silicon would correspond to a DD CRN).

How the sponge phase transforms into the DG crystal is, at present, unknown. The transformation likely involves local translation and rotation of the trihedral mesoatoms to create the correct saddle shapes for the outer terminal surface that can then nest against one another while directing strut orientations along <110> directions. However, the sponge network must be occasionally disrupted in order to split the single non-catenated network with its wide range of loop sizes into two independent, inter-catenated, opposite chirality, 10-3 loop networks. If the 3 principles for mesoatom identification (page 6) are correct, then this network fragmentation and re-fusing of the constituent mesoatoms should obey these principles. Future work needs to address the details of the distributions of the strut-strut angles, the strut lengths and strut directions, as well as characterization of the loop distributions, topologies and various types of network point defects (e.g. f = 4 and 5 nodes as well as network breaks) as the structure evolves across the interface from single disordered sponge network to ordered double gyroid networks.

The detailed structure of the sponge phase as revealed by the 3D SVSEM reconstruction strongly suggests that trihedral mesoatom units are the primordial mesoatomic building blocks in the "sponge-first" pathway to the DG. The "average" primordial mesoatom can be specified using the spread of its characteristic features (i.e. distributions of strut-strut angles, strut lengths, mesoatom volume, IMDS curvature and surface area, dihedral angles between a mesoatom with its linked neighbors as well as the partitioning of its outer terminal surfaces with the surrounding mesoatoms and specification of the type and number of contacting neighbors).

**Learning about Malleable Mesoatoms from Defects**

Studying how the local symmetries of a crystal can be disrupted by various defects, yet still allow the distorted structure to accomodate into the surrounding crystal with an overall small strain field (and hence low energy), can give insight into how mesoatomic units can adapt to their surroundings, which itself reflects the combinations of thermodynamics of the interior chain packing within mesoatoms, as well as the effects of inter-mesoatomic packing. Defect identification and classification in tubular network block copolymer crystals is a relatively recent endeavor [59–61]. Unlike defects in atomic crystals where the structure simply rearranges the immutable atoms, in self assembled crystals with malleable mesoatomic units, distortions and defects can and do strongly alter the shape and symmetry of the basic motif. Since defects disrupt the periodic packing scheme in the crystal; their presence influences both reciprocal space data (scattering) and real space data (microscopy). Therefore, as we have previously discussed, real space analysis is necessary for detailed characterization. Next we do a brief survey of defects with particular attention as to how various defects in the DG phase create changes to the malleable $f$ = 3 DG mesoatomic units.

Defects can be classified as point, line or surface imperfections (0, 1 and 2 D type defects) that respectively break symmetry at a point or along a continuous curve or over a surface. Due to the mesoatoms forming a double inter-catenated network structure, the notion of "point" defects needs to be extended to allow for somewhat larger motifs sometimes containing multiple mesoatomic units that together break symmetry in a local region ("point") but allow rapid return of the structure to its ordered symmetries in adjacent regions. A variety of point defects in the DG phase were identified using SVSEMT [59]. These include node defects (so called $f$ defects), loops and donuts, as well as network-network bridges and network strut-break defects. Fig. 17 shows a few examples of $f$, donut, loop, bridge and break defects that are associated with various departures from the basic $f$ = 3 mesoatom unit that is organized into 10-$3^{10}$ inter-catenated left and right handed chiral loops of the DG crystal. (The notation X-Y$^n$ here denotes a loop consisting of X nodes, each node has a functionality (valence) of Y and there are n consecutive nodes circumventing the loop with this valence). Analysis of the network topology of the DG revealed small closed paths, denoted as "donuts" (e.g. a 5-4,3$^4$) that due to the small diameter of the path were not catenated whereas larger "loop" paths containing multiple f defects were inter-catenated (e.g. 9-4,3$^2$,4,3$^5$). In general, point defects are found to occur in small clusters such that away from the cluster, the structure has returned to the normal crystalline ordered packing. These point defects in network phases exemplify the malleability of the mesoatoms, with the ability of the motifs to adapt their local detailed shapes in order to maintain a smooth, continuous IMDS with relatively mild distortions of the minority and majority domain shapes and thicknesses to minimize the excess chain frustration that the defect(s) creates.

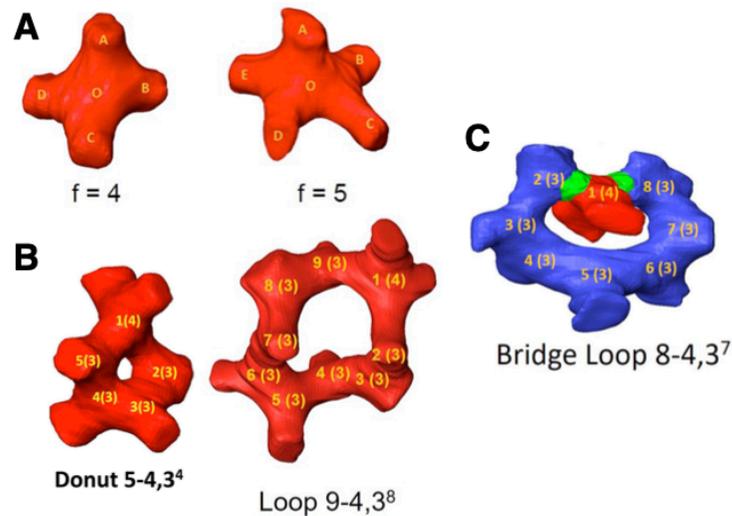

**Fig. 17** Examples of the IMDS within mesoatom point defects occurring in an ordered DG grain. (A) Extracted defect mesoatoms with f = 4 and f = 5. (B) A functionality point defect having one or more extra struts necessitates changes to the network topology and new types of network circuits (donuts and loops) differing from the normal 10-$3^{10}$ loop are formed. (C) A bridge defect occurs when the two networks fuse together. Reproduced from ref. [58]

Dislocations are extended, generally 3D curved, line defects and break the translational symmetry of the crystal. A dislocation is characterized by a unit line vector $\xi$ and a net translational displacement vector (called the Burgers vector) **b**. Dislocations can be classified as pure edge: **b** $\perp$ $\xi$, pure screw: **b** ∥ $\xi$ or mixed [62]. Dislocation defects create an associated strain field that diminishes outwards from their core region with the energy per unit line length scaling with the square of the Burgers vector. A recent TEMT study characterized a dislocation defect in the DG phase [63]. Due to the limited volume of the reconstruction only a relatively short (~ 600 nm) length of dislocation could be investigated. The dislocation line in the reconstruction was of mixed edge and screw character with $\xi$ = [111] and a rather large Burgers vector **b** = $a_o$ [012]. Surprisingly, the core region of the dislocation which normally sustains the greatest distortion, did not exhibit any new mesoatom features, rather a pair of adjacent 5 member and 7 member channel regions were evident instead of the normal 6 member channel arrangement of the networks when viewed along the [111] direction. Away from the dislocation core in the compression side of the defect, the dislocation line apparently created an associated array of bridge type point defects. Undoubtedly, further work will reveal many new types of line defects in tubular networks along with new types of mesoatomic units.

Two dimensional surface defects (i.e. grain boundaries) occur due to impingement of neighboring grains during growth of the ordered phase. Such boundaries are due to the misorientation of the lattices in the neighboring grains and are generally classified as tilt (plane of the boundary is parallel to the axis of misorientation), or twist, plane of the boundary is normal to the axis of rotation) or more generally, of mixed tilt and twist character. A twin boundary is a special type of tilt grain boundary where the boundary acts as a mirror plane for the adjacent grains. Twins are quite common low energy defects in hard matter. Indeed, recently numerous twins have been found in both DG and DD phases [60,61]. Twins in the DG occur on (422) planes and since the two networks are enantiomorphic, the twin acts as a *topological* mirror. The influence of the twin boundary on the mesoatom networks depends if the nodes reside on the boundary or adjacent to the boundary. Three new types of achiral mesoatoms are created on the DG twin boundary as well as two new types of achiral loops. The IMDS is smooth and continuous across the twin boundary with and the new IMDS within the mesoatoms has similar mean and Gaussian curvatures as the normal IMDS within a 16b mesoatom, consistent with a low energy defect.

Twins in the DD occur on (222) planes (see Fig. 18). Interestingly, in atomic diamond (the single network structure of C, Si, Ge etc.) twins occur on (111) planes. For the DD BCP network, the mesoatom network with its nodes offset from the boundary has the same structure as that of a twin in hard diamond where the struts (corresponding to the atomic bonds in diamond) connect nodes on either side of the boundary are perpendicular to the boundary and retain their $T_d$ symmetry. As was the case for the DG twin, the nodes of the second mesoatom network, which lie in the plane of the boundary and as such must exhibit mirror symmetry parallel to the boundary, transform to adopt $D_{3h}$ symmetry (see Fig. 18 B). Twinning results in substantial modification of the normal mesoatom tetrahedral $T_d$ point group symmetry to form two new types of mesoatoms (pentahedral $f$ =5 and trihedral $f$ = 3) which both adopt $D_{3h}$ symmetry. These mesoatoms alternate and link to form a hexagonal mesh comprised of $(6-(5,3)^3)$ loops in the plane of the boundary. Thus, the packing requirement of the mirror defect induces the mesoatomic point group symmetry to change from $T_d$ to $D_{3h}$.

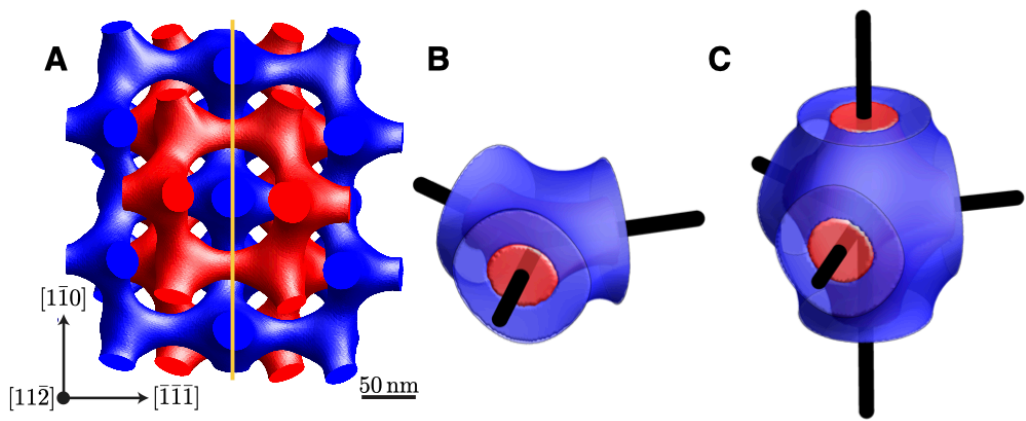

**Fig. 18** A twin boundary defect (mirror plane indicated by the orange line) visualized by 3D SVSEM tomography of a PS-PDMS diblock creates two new types of DD mesoatoms. (A) The (222) twin boundary in the DD phase as viewed along [112] (adapted from [60]). Two new types of mesoatoms having $f$ = 3 (B) and $f$ = 5 (C) are formed on the TB, both with $D_{3h}$ point group symmetry instead of the $T_d$ tetrahedral symmetry of normal DD mesoatoms. Experimental images courtesy of X. Feng and M. Dimitriyev.

An additional key role for malleable mesoatoms lies with their contribution in order – order phase transformations in BCPs [27,64–67]. In order to convert from one phase with a particular space and point group symmetry and mesoatom(s) to a second phase with a different space and point group symmetry mesoatom(s), the mesoatoms must undergo a size and shape transformation. In many studies, the new phase forms an epitaxial relationship with the existing phase that provides a pathway for the transformation. Minimization of the disruptions to preferred packing on either side of the inter-phase boundary often result narrow transition zone across which the mesoatoms undergo restructuring, which implies the existence of new intermediate types of mesoatomic units.

**Concluding Remarks**

We conclude with some brief remarks about basic challenges and questions opened up by the mesoatomic concept when extended to chain architectures beyond the basic "polymer amphiphile" shape of linear AB diblock copolymers. Our simple mesoatom is defined by its inner terminal surface, the IMDS and its outer terminal surface. Such discrete mesoatoms aggregate and pack via brush-brush interactions across the exposed terminal surfaces as well as linking and fusing of nearly parallel blocks across strut faces to smoothly extend the IMDS.  Clearly, this mesoatom concept readily applies to tubular network forming AB diblock – homopolymer A or B blends—appropriately generalized to incorporate "guest" homopolymers in either the tubular or matrix domains.  However, definition of the mesoatom of AB diblock double-network does not simply generalize to double networks formed by ABA triblocks (e.g. [68,69]), even of the same symmetry, since a portion of chain configurations bridge from one tubular network to the other spanning the mid-block matrix  [70] .  For example in DG, to dissect out the 16b mesoatoms (as argued for AB diblocks) from the final structure requires cutting bridging B chains to form the outer terminal mesoatom surface.  In general, the presence of bridging blocks in the matrix phase that covalently connect two different IMDS would then require the choice of a mesoatom with two IMDS (i.e. mesoatom 16a in Fig. 15B).  A three sub-domain, two IMDS mesoatom version of 16a would work for alternating ABC gyroid predicted and observed in linear ABC terblocks  [71,72]. Moreover, in practice for terpolymers with three solvent-block interactions, it is very likely that as solvent evaporates during the assembly, the shape and symmetry of the primordial mesoatoms evolve due to variations in the relative strength of enthalpic and entropic interactions as well as relative component volume fractions. Thus, the primordial dilute solution mesoatoms will likely evolve as the (necessarily preferential for 3 blocks) solvent evaporates, causing, for example, a primordial mesoatom initially comprised of two regions and one IMDS (say an A domain + solvent core region and an outer  mixed B - C + higher solvent content region) to evolve during aggregation to demix the B & C blocks as solvent evaporates to create a new second IMDS between B and C as well as enabling the A and C blocks to link up to form tubular networks.

Beyond linear architectures, much more complex polycontinuous network topologies are predicted for ABC miktoarm stars, including extended and linked lines of periodically spaced triple junctions where all three domains meet  [73,74].  Here the constraint for all 3 blocks to covalently link at a single junction creates a new type of IMDS where the junctions are confined to parallel lines as opposed to spreading uniformly over surfaces [75].  Mesoatoms for star architectures will likely internally partition to reflect the relative volume fractions of each component and the cost of the various types of IMDS between pairs of blocks, predicted to lead complex patterns of interwinding network domains, such as "striped gyroids" [74]. Whether a single, generic set of rules can be constructed to divine mesoatomic shapes when accounting for the vast variations of non-linear molecular architectures, inter-domain topologies and crystallographic (and potentially even quasi-crystallographic) symmetries remains as a challenge. Clearly, the rich and ever-expanding palette of supramolecular chemistry demands improved understanding of how the molecules manage their local environments along the way from either the initial melt state or from the dilute solution state to the final ever-expanding suite of ordered morphologies.

**Acknowledgements**

The authors are grateful to M. Dimitriyev, C. Burke and W. Shan for stimulating discussions and valuable comments on this work, as well as M. Dimitriyev, X. Feng and W. Shan for contributing figure elements.   This research was supported by the US Department of Energy (DOE), Office of Basic Energy Sciences, Division of Materials Sciences and Engineering under award DE-SC0022229.

**References**

[1]   S. Hyde, B. W. . Ninham, and S. Andersson, *The Language of Shape: The Role of Curvature in Condensed Matter Physics, Chemistry, and Biology* (Elsevier, Amsterdam, 2007).
[2]   F. S. Bates and G. H. Fredrickson, *Block Copolymer Thermodynamics: Theory and Experiment*, Annu. Rev. Phys. Chem. **41**, 525 (1990).
[3]   Z. Su, R. Zhang, X.-Y. Yan, Q.-Y. Guo, J. Huang, W. Shan, Y. Liu, T. Liu, M. Huang, and S. Z. D. Cheng, *The Role of Architectural Engineering in Macromolecular Self-Assemblies via Non-Covalent Interactions: A Molecular LEGO Approach*, Progress in Polymer Science **103**, 101230 (2020).
[4]   J. N. Israelachvili, *Soft and Biological Structures*, in *Intermolecular and Surface Forces*, 3rd ed (Academic Press, Burlington, MA, 2011), pp. 535–576.
[5]   E. E. Dormidontova and T. P. Lodge, *The Order−Disorder Transition and the Disordered Micelle Regime in Sphere-Forming Block Copolymer Melts*, Macromolecules **34**, 9143 (2001).
[6]   G. M. Grason, *Ordered Phases of Diblock Copolymers in Selective Solvent*, The Journal of Chemical Physics **126**, 114904 (2007).
[7]   D. J. Mitchell, G. J. T. Tiddy, L. Waring, T. Bostock, and M. P. McDonald, *Phase Behaviour of Polyoxyethylene Surfactants with Water. Mesophase Structures and Partial Miscibility (Cloud Points)*, J. Chem. Soc., Faraday Trans. 1 **79**, 975 (1983).
[8]   P. Sakya, J. M. Seddon, R. H. Templer, R. J. Mirkin, and G. J. T. Tiddy, *Micellar Cubic Phases and Their Structural Relationships: The Nonionic Surfactant System $C_{12}EO_{12}$/Water*, Langmuir **13**, 3706 (1997).
[9]   T. P. Lodge, B. Pudil, and K. J. Hanley, *The Full Phase Behavior for Block Copolymers in Solvents of Varying Selectivity*, Macromolecules **35**, 4707 (2002).
[10]  E. L. Thomas, D. J. Kinning, D. B. Alward, and C. S. Henkee, *Ordered Packing Arrangements of Spherical Micelles of Diblock Copolymers in Two and Three Dimensions*, Macromolecules **20**, 2934 (1987).

[11] J. Charvolin and J. F. Sadoc, *Periodic Systems of Frustrated Fluid Films and « micellar » Cubic Structures in Liquid Crystals*, J. Phys. France **49**, 521 (1988).

[12] G. Grason, *The Packing of Soft Materials: Molecular Asymmetry, Geometric Frustration and Optimal Lattices in Block Copolymer Melts*, Physics Reports **433**, 1 (2006).

[13] S. Lee, C. Leighton, and F. S. Bates, *Sphericity and Symmetry Breaking in the Formation of Frank–Kasper Phases from One Component Materials*, Proc. Natl. Acad. Sci. U.S.A. **111**, 17723 (2014).

[14] M. Huang et al., *Selective Assemblies of Giant Tetrahedra via Precisely Controlled Positional Interactions*, Science **348**, 424 (2015).

[15] G. M. Grason, B. A. DiDonna, and R. D. Kamien, *Geometric Theory of Diblock Copolymer Phases*, Phys. Rev. Lett. **91**, 058304 (2003).

[16] A.-C. Shi, *Frustration in Block Copolymer Assemblies*, J. Phys.: Condens. Matter **33**, 253001 (2021).

[17] K. D. Dorfman, *Frank–Kasper Phases in Block Polymers*, Macromolecules **54**, 10251 (2021).

[18] Y. Liu et al., *Mesoatom Alloys via Self-Sorting Approach of Giant Molecules Blends*, Giant **4**, 100031 (2020).

[19] A. Reddy, M. B. Buckley, A. Arora, F. S. Bates, K. D. Dorfman, and G. M. Grason, *Stable Frank–Kasper Phases of Self-Assembled, Soft Matter Spheres*, Proc. Natl. Acad. Sci. U.S.A. **115**, 10233 (2018).

[20] J. Charvolin and J. F. Sadoc, *Periodic Systems of Frustrated Fluid Films and « bicontinuous » Cubic Structures in Liquid Crystals*, J. Phys. France **48**, 1559 (1987).

[21] G. E. Schröder-Turk, A. Fogden, and S. T. Hyde, *Bicontinuous Geometries and Molecular Self-Assembly: Comparison of Local Curvature and Global Packing Variations in Genus-Three Cubic, Tetragonal and Rhombohedral Surfaces*, Eur. Phys. J. B **54**, 509 (2006).

[22] E. L. Thomas, D. M. Anderson, C. S. Henkee, and D. Hoffman, *Periodic Area-Minimizing Surfaces in Block Copolymers*, Nature **334**, 598 (1988).

[23] M. W. Matsen, *The Standard Gaussian Model for Block Copolymer Melts*, J. Phys.: Condens. Matter **14**, R21 (2002).

[24] A. Reddy, X. Feng, E. L. Thomas, and G. M. Grason, *Block Copolymers beneath the Surface: Measuring and Modeling Complex Morphology at the Subdomain Scale*, Macromolecules **54**, 9223 (2021).

[25] M. J. Park, K. Char, J. Bang, and T. P. Lodge, *Order−Disorder Transition and Critical Micelle Temperature in Concentrated Block Copolymer Solutions*, Macromolecules **38**, 2449 (2005).

[26] K. Kim, M. W. Schulze, A. Arora, R. M. Lewis, M. A. Hillmyer, K. D. Dorfman, and F. S. Bates, *Thermal Processing of Diblock Copolymer Melts Mimics Metallurgy*, Science **356**, 520 (2017).

[27] D. A. Hajduk, P. E. Harper, S. M. Gruner, C. C. Honeker, G. Kim, E. L. Thomas, and L. J. Fetters, *The Gyroid: A New Equilibrium Morphology in Weakly Segregated Diblock Copolymers*, Macromolecules **27**, 4063 (1994).

[28] M. Schick, *Avatars of the Gyroid*, Physica A: Statistical Mechanics and Its Applications **251**, 1 (1998).

[29] I. Prasad, H. Jinnai, R.-M. Ho, E. L. Thomas, and G. M. Grason, *Anatomy of Triply-Periodic Network Assemblies: Characterizing Skeletal and Inter-Domain Surface Geometry of Block Copolymer Gyroids*, Soft Matter **14**, 3612 (2018).

[30] A. Reddy, M. S. Dimitriyev, and G. M. Grason, *Medial Packing and Elastic Asymmetry Stabilize the Double-Gyroid in Block Copolymers*, Nat Commun **13**, 2629 (2022).

[31] G. E. Schröder-Turk, A. Fogden, and S. T. Hyde, *Local v/a Variations as a Measure of Structural Packing Frustration in Bicontinuous Mesophases, and Geometric Arguments for

*an Alternating ${\rm Im}\overline{{\mathsf 3}}{\rm M}$ (I-WP) Phase in Block-Copolymers with Polydispersity*, Eur. Phys. J. B **59**, 115 (2007).

[32] A. H. Schoen, *Reflections Concerning Triply-Periodic Minimal Surfaces*, Interface Focus. **2**, 658 (2012).
[33] P. D. Olmsted and S. T. Milner, *Strong Segregation Theory of Bicontinuous Phases in Block Copolymers*, Macromolecules **31**, 4011 (1998).
[34] P. J. F. Gandy, S. Bardhan, A. L. Mackay, and J. Klinowski, *Nodal Surface Approximations to the P,G,D and I-WP Triply Periodic Minimal Surfaces*, Chemical Physics Letters **336**, 187 (2001).
[35] M. Wohlgemuth, N. Yufa, J. Hoffman, and E. L. Thomas, *Triply Periodic Bicontinuous Cubic Microdomain Morphologies by Symmetries*, Macromolecules **34**, 6083 (2001).
[36] A. F. Wells, *Three-Dimensional Nets and Polyhedra* (Wiley, New York, 1977).
[37] X. Feng et al., *Seeing Mesoatomic Distortions in Soft-Matter Crystals of a Double-Gyroid Block Copolymer*, Nature **575**, 175 (2019).
[38] P. F. Damasceno, M. Engel, and S. C. Glotzer, *Predictive Self-Assembly of Polyhedra into Complex Structures*, Science **337**, 453 (2012).
[39] E. S. Harper, G. van Anders, and S. C. Glotzer, *The Entropic Bond in Colloidal Crystals*, Proc. Natl. Acad. Sci. U.S.A. **116**, 16703 (2019).
[40] E. Bianchi, J. Largo, P. Tartaglia, E. Zaccarelli, and F. Sciortino, *Phase Diagram of Patchy Colloids: Towards Empty Liquids*, Phys. Rev. Lett. **97**, 168301 (2006).
[41] F. Sciortino and E. Zaccarelli, *Reversible Gels of Patchy Particles*, Current Opinion in Solid State and Materials Science **15**, 246 (2011).
[42] F. Smallenburg, L. Filion, and F. Sciortino, *Erasing No-Man's Land by Thermodynamically Stabilizing the Liquid–Liquid Transition in Tetrahedral Particles*, Nature Phys **10**, 653 (2014).
[43] N. Kern and D. Frenkel, *Fluid–Fluid Coexistence in Colloidal Systems with Short-Ranged Strongly Directional Attraction*, The Journal of Chemical Physics **118**, 9882 (2003).
[44] A. Neophytou, D. Chakrabarti, and F. Sciortino, *Facile Self-Assembly of Colloidal Diamond from Tetrahedral Patchy Particles via Ring Selection*, Proc. Natl. Acad. Sci. U.S.A. **118**, e2109776118 (2021).
[45] A. Kumar and V. Molinero, *Why Is Gyroid More Difficult to Nucleate from Disordered Liquids than Lamellar and Hexagonal Mesophases?*, J. Phys. Chem. B **122**, 4758 (2018).
[46] M. Marriott, L. Lupi, A. Kumar, and V. Molinero, *Following the Nucleation Pathway from Disordered Liquid to Gyroid Mesophase*, J. Chem. Phys. **150**, 164902 (2019).
[47] A. J. Mukhtyar and F. A. Escobedo, *Developing Local Order Parameters for Order–Disorder Transitions From Particles to Block Copolymers: Methodological Framework*, Macromolecules **51**, 9769 (2018).
[48] D. Morphew, J. Shaw, C. Avins, and D. Chakrabarti, *Programming Hierarchical Self-Assembly of Patchy Particles into Colloidal Crystals* via *Colloidal Molecules*, ACS Nano **12**, 2355 (2018).
[49] A. B. Rao, J. Shaw, A. Neophytou, D. Morphew, F. Sciortino, R. L. Johnston, and D. Chakrabarti, *Leveraging Hierarchical Self-Assembly Pathways for Realizing Colloidal Photonic Crystals*, ACS Nano **14**, 5348 (2020).
[50] Y. Talmon, *Cryo-Electron Microscopy of Vitrified Aqueous Specimens*, Ultramicroscopy **17**, 167 (1985).
[51] Y. Talmon, *The Study of Nanostructured Liquids by Cryogenic-Temperature Electron Microscopy — A Status Report*, Journal of Molecular Liquids **210**, 2 (2015).
[52] A. L. Parry, P. H. H. Bomans, S. J. Holder, N. A. J. M. Sommerdijk, and S. C. G. Biagini, *Cryo Electron Tomography Reveals Confined Complex Morphologies of Tripeptide-Containing Amphiphilic Double-Comb Diblock Copolymers*, Angew. Chem. Int. Ed. **47**, 8859 (2008).

[53] G. Porte, J. Marignan, P. Bassereau, and R. May, *Shape Transformations of the Aggregates in Dilute Surfactant Solutions : A Small-Angle Neutron Scattering Study*, J. Phys. France **49**, 511 (1988).

[54] Y.-Y. Won, A. K. Brannan, H. T. Davis, and F. S. Bates, *Cryogenic Transmission Electron Microscopy (Cryo-TEM) of Micelles and Vesicles Formed in Water by Poly(Ethylene Oxide)-Based Block Copolymers*, J. Phys. Chem. B **106**, 3354 (2002).

[55] S. Jain and F. S. Bates, *On the Origins of Morphological Complexity in Block Copolymer Surfactants*, Science **300**, 460 (2003).

[56] G. M. Scheutz, M. A. Touve, A. S. Carlini, J. B. Garrison, K. Gnanasekaran, B. S. Sumerlin, and N. C. Gianneschi, *Probing Thermoresponsive Polymerization-Induced Self-Assembly with Variable-Temperature Liquid-Cell Transmission Electron Microscopy*, Matter **4**, 722 (2021).

[57] J. Korpanty, L. R. Parent, N. Hampu, S. Weigand, and N. C. Gianneschi, *Thermoresponsive Polymer Assemblies via Variable Temperature Liquid-Phase Transmission Electron Microscopy and Small Angle X-Ray Scattering*, Nat Commun **12**, 6568 (2021).

[58] H. Jinnai, H. Hasegawa, Y. Nishikawa, G. J. A. Sevink, M. B. Braunfeld, D. A. Agard, and R. J. Spontak, *3D Nanometer-Scale Study of Coexisting Bicontinuous Morphologies in a Block Copolymer/Homopolymer Blend*, Macromol. Rapid Commun. **27**, 1424 (2006).

[59] X. Feng, H. Guo, and E. L. Thomas, *Topological Defects in Tubular Network Block Copolymers*, Polymer **168**, 44 (2019).

[60] X. Feng, M. Zhuo, H. Guo, and E. L. Thomas, *Visualizing the Double-Gyroid Twin*, Proc. Natl. Acad. Sci. U.S.A. **118**, e2018977118 (2021).

[61] X. Feng, M. S. Dimitriyev, and E. L. Thomas, *Soft, Malleable Double Diamond Twin*, Proceedings of the National Academy of Sciences (in Press) (2022).

[62] J. P. Hirth and J. Lothe, *Theory of Dislocations*, 2nd ed (Krieger Pub. Co, Malabar, FL, 1992).

[63] T. Miyata, H.-F. Wang, T. Suenaga, D. Watanabe, H. Marubayashi, and H. Jinnai, *Dislocation-Induced Defect Formation in a Double-Gyroid Network*, Macromolecules **55**, 8143 (2022).

[64] D. A. Hajduk, R.-M. Ho, M. A. Hillmyer, F. S. Bates, and K. Almdal, *Transition Mechanisms for Complex Ordered Phases in Block Copolymer Melts*, J. Phys. Chem. B **102**, 1356 (1998).

[65] M. W. Matsen, *Cylinder ↔ Gyroid Epitaxial Transitions in Complex Polymeric Liquids*, Phys. Rev. Lett. **80**, 4470 (1998).

[66] C.-Y. Wang and T. P. Lodge, *Kinetics and Mechanisms for the Cylinder-to-Gyroid Transition in a Block Copolymer Solution*, Macromolecules **35**, 6997 (2002).

[67] J. Jung, J. Lee, H.-W. Park, T. Chang, H. Sugimori, and H. Jinnai, *Epitaxial Phase Transition between Double Gyroid and Cylinder Phase in Diblock Copolymer Thin Film*, Macromolecules **47**, 8761 (2014).

[68] A. Avgeropoulos, B. J. Dair, N. Hadjichristidis, and E. L. Thomas, *Tricontinuous Double Gyroid Cubic Phase in Triblock Copolymers of the ABA Type*, Macromolecules **30**, 5634 (1997).

[69] H. Jinnai, Y. Nishikawa, R. J. Spontak, S. D. Smith, D. A. Agard, and T. Hashimoto, *Direct Measurement of Interfacial Curvature Distributions in a Bicontinuous Block Copolymer Morphology*, Phys. Rev. Lett. **84**, 518 (2000).

[70] M. W. Matsen and R. B. Thompson, *Equilibrium Behavior of Symmetric ABA Triblock Copolymer Melts*, The Journal of Chemical Physics **111**, 7139 (1999).

[71] T. H. Epps, E. W. Cochran, T. S. Bailey, R. S. Waletzko, C. M. Hardy, and F. S. Bates, *Ordered Network Phases in Linear Poly(Isoprene- b -Styrene- b -Ethylene Oxide) Triblock Copolymers*, Macromolecules **37**, 8325 (2004).

[72] J. Qin, F. S. Bates, and D. C. Morse, *Phase Behavior of Nonfrustrated ABC Triblock Copolymers: Weak and Intermediate Segregation*, Macromolecules **43**, 5128 (2010).
[73] M. G. Fischer, L. de Campo, J. J. K. Kirkensgaard, S. T. Hyde, and G. E. Schröder-Turk, *The Tricontinuous 3* Ths *(5) Phase: A New Morphology in Copolymer Melts*, Macromolecules **47**, 7424 (2014).
[74] J. J. K. Kirkensgaard, M. E. Evans, L. de Campo, and S. T. Hyde, *Hierarchical Self-Assembly of a Striped Gyroid Formed by Threaded Chiral Mesoscale Networks*, Proc. Natl. Acad. Sci. U.S.A. **111**, 1271 (2014).
[75] S. Okamoto, H. Hasegawa, T. Hashimoto, T. Fujimoto, H. Zhang, T. Kazama, A. Takano, and Y. Isono, *Morphology of Model Three-Component Three-Arm Star-Shaped Copolymers*, Polymer **38**, 5275 (1997).

**Supporting Material**

**Supporting Movie 1**
(http://www.pse.umass.edu/sites/default/files/grason/images/dg_89_meso_skel.gif) – Sequence of first $N$ = 89 mesoatom additions in simulated growth of DG crystal (solid mesoatoms on left and corresponding occupied skeletons on right)